\documentclass[twocolumn]{jpsj3}
\title{%
Quasiclassical Theory of the Josephson Effect
in Ballistic Graphene Junctions
}

\author{%
Yositake {\sc Takane} and Ken-Ichiro {\sc Imura}
}

\inst{%
Department of Quantum Matter, Graduate School of Advanced Sciences of Matter,\\
Hiroshima University, Higashihiroshima, Hiroshima 739-8530, Japan
}

\recdate{ \hspace{50mm} }

\abst{%
The stationary Josephson effect in a system of ballistic graphene is studied
in the framework of quasiclassical Green's function theory.
Reflecting the ultimate two-dimensionality of graphene,
a Josephson junction involving a graphene sheet embodies
what we call a planar Josephson junction, in which superconducting electrodes
partially cover the two-dimensional graphene layer,
achieving a planar contact with it.
For capturing this feature we employ a model of tunneling Hamiltonian
that also takes account of the effects of inhomogeneous carrier density.
Within the effective mass approximation we derive a general formula
for the Josephson current, revealing characteristic features of
the superconducting proximity effect in the planar Josephson junction.
The same type of analysis has been equally applied to mono-,
bi- and arbitrary $N$-layer cases.
}

\kword{%
Josephson current, multilayer graphene, quasiclassical Green's function
}

\begin{document}
\sloppy
\maketitle

\section{Introduction}

When two superconductors are placed spatially apart but weakly coupled,
a finite amount of dissipationless equilibrium current is generally
induced between the two superconductors.
Though this effect, the stationary Josephson effect,
was originally predicted~\cite{josephson} for a system of tunneling junction,
i.e., two superconductors separated by a thin insulating barrier,
the same effect is now known to exist in a broad range of
superconducting junctions, and especially in the ones mediated by
various other non-superconducting elements, such as
a normal metal,~\cite{likharev}
a two-dimensional electron gas,~\cite{van-wees}
or a quantum dot.~\cite{Martin-Rodero}
Yet, the characteristic behavior of the dissipationless current
in such Josephson junctions is strongly influenced by the electronic property
of the element inserted between the two superconductors
and by the way that element is placed between them.

During the last several years, a variant of such Josephson junctions
realized in a system of graphene,
i.e., the superconductor-graphene-superconductor (SGS) junction,
has become a target of
intense theoretical~\cite{wakabayashi,titov,moghaddam,gonzalez,
black-schaffer,hayashi,hagymasi}
and experimental~\cite{heersche,sato,du,ojeda,kanda,tomori} studies.
Much has been studied on how the Josephson current is affected by
the unique band structure of a graphene monolayer,~\cite{novoselov,castro_neto}
in which the conduction and valence bands touch conically
at $K_{+}$ and $K_{-}$ points in the Brillouin zone (the Dirac points).
An interesting result is that
in the SGS Josephson junctions involving a monolayer of graphene,
the critical current $I_{\rm c}$ at zero temperature remains finite
at $\epsilon = 0$ (when the chemical potential is placed at the level of
the Dirac points) in spite of the vanishing density of states.~\cite{titov}

On contrary, most of the theoretical studies performed so far have neglected
the unique structural character of the SGS Josephson junction.
Since graphene is an isolated ideal two-dimensional electron system,
a natural way to fabricate an SGS junction is
to deposit superconducting electrodes on top of
a graphene flake.~\cite{heersche,sato,du,ojeda,kanda,tomori}
Then, the graphene sheet beneath the superconducting electrodes 
acquires a two-dimensional, planar contact with the electrodes.
This is indeed a very unique situation, opposing,
e.g., to the case of a two-dimensional electron gas imbedded in
a semiconductor hetero-structure that has a one-dimensional (linear) contact
with superconducting electrodes.~\cite{van-wees}
In most of the existing theoretical studies~\cite{titov,moghaddam,
black-schaffer,hagymasi} the planar character of the SGS junction
is poorly taken into account.
It is usually assumed that
an energy-independent effective pair potential $\Delta_{\rm eff}$ 
is induced inside the graphene sheet
in the region covered by the superconductors.
But this very assumption reduces the SGS junction of {\it intrinsically}
planar nature to a conventional linear junction.
The latter consists of a graphene sheet
of a finite length placed between two (hypothetical) graphene superconductors.

The planar character of the SGS junction manifests
in the temperature ($T$-) dependence of the Josephson current.
The conventional formulation based on
the energy-independent $\Delta_{\rm eff}$ fails, by its construction,
to describe this feature.
Of course, one can always
{\it assume} a $T$-dependence of $\Delta_{\rm eff} = \Delta_{\rm eff}(T)$
and discuss the $T$-dependence of Josephson current, but this is not more than
an {\it ad hoc} solution.
For example, the $T$-dependence of the Josephson current has been calculated
by assuming $\Delta_{\rm eff} (T)$
as described by the BCS theory.~\cite{hagymasi}
Here, in this paper, extending an earlier work of ref.~\citen{takane1},
we give a more fundamental solution to this issue, allowing for a reliable
prediction of the $T$-dependence of Josephson current
in the planar SGS junction.
In the approach undertaken we explicitly take account of the tunneling of
electrons in the planar regions of contact.
In other words, we describe the coupling between the graphene sheet
and the superconducting electrodes by a tunneling Hamiltonian.
The strength of the tunnel coupling is controlled by the parameter $\Gamma$.

Another advantage of our approach, which we aim at reporting in this paper,
is that it allows for a systematic generalization to
the case of multi-layer graphene.
A renewed insight into the use of quasiclassical Green's
function~\cite{eilenberger,larkin} in the system of SGS junction enables this.
Here, the electronic property of mono-, bi- and multi-layer graphenes
is treated under the effective mass approximation.
On one hand, most of the published theoretical results have focused on
the case of monolayer graphene, with the exception of ref.~\citen{hayashi}
treating the case of bilayer.
The experiments are, on the other hand,
not restricted to the case of monolayer.~\cite{heersche,sato,kanda}
The bilayer and multi-layer ($N$-layer with $N=3,4,\cdots$) cases are
(and will be in the near future) equally highlighted experimentally.
A bilayer graphene has a nearly quadratic energy dispersion.~\cite{maccann}
In the case of $N$-layer graphene with $N \ge 3$,
mono- and bi-layer type dispersions coexist,
but a basic tendency is determined only by the parity of $N$.~\cite{koshino}
We show that the Josephson effect in the SGS junction also exhibits
such an even-odd feature with respect to $N$, the number of layers.

The backbone of our approach is the use of a tunneling Hamiltonian
at the level of the modeling (of planar structure),
and of the quasiclassical Green's function approach
on which subsequent studies of the Josephson current are entirely based.
In addition to the full formulation of the present approach
that has not been done before,
the new ingredients first taken into account in this paper are the following.
The effect of inhomogeneous carrier density:
due to the contact with superconductors, the carrier density in
the graphene sheet becomes higher in the region covered by superconductors.
Since this results in a mismatch in the Fermi wave number at the interface
between the covered and uncovered regions,
we expect a reduction of the Josephson current.
This effect was ignored in ref.~\citen{takane1}.
As mentioned earlier,
we also first deal with the case of an arbitrary number $N$ of layers.
Taking these two new elements into account,
we study the $T$- and $\Gamma$-dependences of the Josephson current.
To avoid unnecessary complication
the chemical potential $\mu$ is assumed to be away from the Dirac point,
and we consider only the ballistic regime.

The paper is arranged as follows.
In \S2, we start by modeling the SGS planar Josephson junction
considering generally the case of $N$-layer graphene.
We then introduce a thermal Green's function adapted for the system
in consideration.
In \S3, we derive a general formula for the Josephson current
in the cases of mono- and bi-layer within the quasiclassical approximation
applied with the use of the thermal Green's function introduced in \S2.
In \S4, we extend the argument in the previous section
to the case of arbitrary $N$-layer.
In \S5, the behavior of the Josephson critical current is studied
in the short-junction limit.
We discuss analytic expressions obtained in some specific cases of parameters.
Otherwise, the critical current is evaluated numerically.
Implications of the obtained results are discussed
in the light of current experimental studies that focus on the same limit.
Section 6 is devoted to summary.
We set $k_{\rm B} = \hbar = 1$ throughout the paper.

\section{Modeling of the SGS Junction}

We start by modeling the SGS junction, taking properly into account 
both its structural uniqueness and the nature of electronic states
in the intermediate non-superconducting element,
i.e., in the graphene sheet.
Here, we consider the case of a graphene sheet consisting of
generally $N$ layers (treating the cases of mono-, bi-, and
multi-layer graphene in parallel),
which we describe in the effective mass approximation.
As a basic theoretical tool indispensable for the subsequent analyses,
we then introduce a thermal Green's function adapted for the system
of SGS junction incorporating an $N$-layer graphene sheet.
The influence of superconducting electrodes on quasiparticles in graphene
is taken into account by a self-energy associated with that Green's function.

\begin{figure}[btp]
\begin{center}
\includegraphics[height=4.5cm]{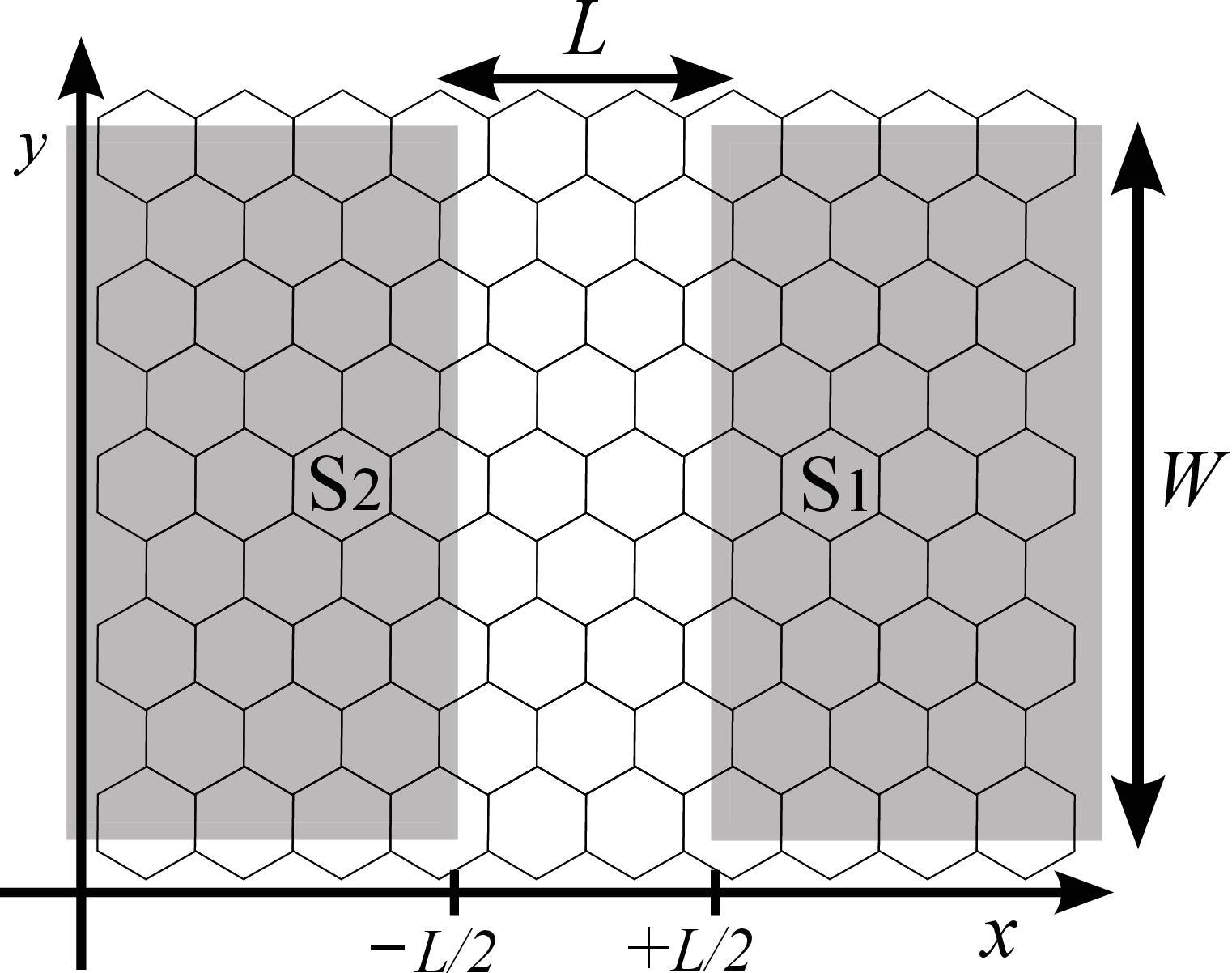}
\end{center}
\caption{Planar junction geometry: Josephson junction consisting of a graphene
sheet on which two superconductors ${\rm S}_{1}$ and ${\rm S}_{2}$ of
width $W$ are deposited with separation $L$.
The pair potential is assumed to be $\Delta {\rm e}^{{\rm i}\varphi/2}$
in ${\rm S}_{1}$ and $\Delta {\rm e}^{-{\rm i}\varphi/2}$ in ${\rm S}_{2}$.
}
\end{figure}
The SGS junction we consider has a construction as depicted in Fig.~1.
Two superconductors ${\rm S}_{1}$ and ${\rm S}_{2}$ (of width $W$)
are placed (with separation $L$)
on top of a clean graphene sheet,
where ${\rm S}_{1}$ and ${\rm S}_{2}$ occupy the region of $L/2 \le x$
and that of $x \le -L/2$, respectively.
Note that only the top (first) layer is in contact with
${\rm S}_{1}$ and ${\rm S}_{2}$.
We assume that the pair potential in ${\rm S}_{1}$ and ${\rm S}_{2}$
is given by
\begin{align}
  \Delta (x)
  = \left\{ \begin{array}{cc}
               \Delta {\rm e}^{{\rm i}\varphi/2}
               & (L/2 < x) \\
               0 & (-L/2 < x < L/2) \\
               \Delta {\rm e}^{-{\rm i}\varphi/2}
               & (x < -L/2)
            \end{array}
    \right. .
\end{align}
Let us assume that
the coupling of the graphene sheet and the superconductors
is described by a tunneling Hamiltonian.
Then, the resulting proximity effect on quasiparticles in graphene
is described by the self-energy~\cite{mcmillan}
for the thermal Green's function as given below.

The coupling with ${\rm S}_{1}$ and ${\rm S}_{2}$ also
induces carrier doping in the graphene sheet, i.e.,
the carrier density in the covered region of $|x| > L/2$ becomes higher
than that in the uncovered region of $|x| < L/2$.
This effect cannot be described by the self-energy,
so we take this into account by adding the effective potential
of a negative value $-U$ only in the covered region but for every layer.
This potential rapidly varies across the interface at $x = \pm L/2$
with a characteristic length scale
which is comparable to, or shorter than, the Fermi wave length
but longer than the lattice constant $a$ of graphene.~\cite{comment1}
It is convenient to introduce the renormalized chemical potential
$\tilde{\mu}$ defined by
\begin{align}
  \tilde{\mu}
  = \left\{ \begin{array}{cc}
               \mu & (-L/2 < x < L/2) \\
               \mu + U  & (L/2 < |x|)
            \end{array}
    \right. .
\end{align}

Let us turn our attention to the electronic property of the graphene sheet,
which we assume to be composed of $N$ layers.
The unit cell of the hexagonal lattice of monolayer graphene
contains A and B sites.
We assume that graphene layers are subjected to the AB stacking.
That is, A sites in the first layer are located just above B sites
in the second layer, and B sites in the second layer are located just above
A sites in the third layer, and so forth.
To describe electron states in an $N$-layer graphene sheet,
we first employ a tight-binding model on the AB stacked hexagonal lattice
with the nearest-neighbor in-plane transfer integral $\gamma_{0}$ and
the nearest-neighbor vertical coupling
$\gamma_{1}$.~\cite{wallace,mcclure,slonczewski}
These parameters are estimated as $\gamma_{0} \approx 2.8$ eV
and $\gamma_{1} \approx 0.4$ eV.~\cite{castro_neto}

Let $\psi_{{\rm A}l}(\mib{R}_{\rm A})$ ($\psi_{{\rm B}l}(\mib{R}_{\rm B})$)
be the amplitude of the wave function at $\mib{R}_{\rm A}$ ($\mib{R}_{\rm B}$)
on the $l$th layer ($1 \le l \le N$),
where $\mib{R}_{\rm A}$ ($\mib{R}_{\rm B}$) represents
the coordinate of an arbitrary site which belongs to the A (B) sublattice.
Low energy states appear near the $K_{+}$ and $K_{-}$ points
in the two-dimensional Brillouin zone.
The wave vector corresponding to the $K_{\pm}$ point is given by
$\pm \mib{K} = \pm(2\pi/a)(2/3,0)$.
We express $\psi_{{\rm A}l}(\mib{R}_{\rm A})$ and
$\psi_{{\rm B}l}(\mib{R}_{\rm B})$ as
\begin{align}
 \psi_{{\rm A}l}(\mib{R}_{\rm A})
  & =  {\rm e}^{{\rm i}\mib{K}\cdot \mib{R}_{\rm A}}
       F_{{\rm A}l}^{+}(\mib{R}_{\rm A})
     + (-1)^{l-1}{\rm e}^{-{\rm i}\mib{K}\cdot \mib{R}_{\rm A}}
       F_{{\rm A}l}^{-}(\mib{R}_{\rm A}) ,
    \\
 \psi_{{\rm B}l}(\mib{R}_{\rm B})
  & =  {\rm e}^{{\rm i}\mib{K}\cdot \mib{R}_{\rm B}}
       F_{{\rm B}l}^{+}(\mib{R}_{\rm B})
     + (-1)^{l}{\rm e}^{-{\rm i}\mib{K}\cdot \mib{R}_{\rm B}}
       F_{{\rm B}l}^{-}(\mib{R}_{\rm B}) ,
\end{align}
where $F_{{\rm A}l}^{\pm}(\mib{r})$ and $F_{{\rm B}l}^{\pm}(\mib{r})$
are envelop functions.
Let us define the electron state $|\Psi_{\pm}\rangle_{e}$
near the $K_{\pm}$ point by
\begin{align}
  |\Psi_{\pm}\rangle_{e}
  = \sum_{l =1}^{N} F_{{\rm A}l}^{\pm}(\mib{r})|{\rm A}_{l}\rangle_{e}
    + \sum_{l =1}^{N} F_{{\rm B}l}^{\pm}(\mib{r})|{\rm B}_{l}\rangle_{e} ,
\end{align}
and its time-reversed hole state by
\begin{align}
  |\Psi_{\pm}\rangle_{h}
 & = \sum_{l=1}^{N} (-1)^{l-1}
                   F_{{\rm A}l}^{\mp}(\mib{r})^{*}|{\rm A}_{l}\rangle_{h}
      \nonumber \\
 & \hspace{18mm}
    + \sum_{l=1}^{N} (-1)^{l}
                   F_{{\rm B}l}^{\mp}(\mib{r})^{*}|{\rm B}_{l}\rangle_{h} .
\end{align}
Here $|{\rm A}_{l}\rangle_{e}$ ($|{\rm B}_{l}\rangle_{e}$)
represents the basis vector for electron states
on the A (B) sublattice of layer $l$,
and $|{\rm A}_{l}\rangle_{h}$ ($|{\rm B}_{l}\rangle_{h}$)
represents that for hole states.

Within the effective mass approximation,
one can show that these states satisfy
$\bar{H}_{\pm}|\Psi_{\pm}\rangle_{e} = \epsilon |\Psi_{\pm}\rangle_{e}$
and $\bar{H}_{\pm}|\Psi_{\pm}\rangle_{h}
= -\epsilon |\Psi_{\pm}\rangle_{h}$, respectively,
where $\bar{H}_{\pm}$ is an effective Hamiltonian given below~\cite{koshino}
and $\epsilon$ is the energy measured from $\mu$.
If the following basis set
$|{\rm A}_{1}\rangle_{p}$, $|{\rm B}_{1}\rangle_{p}$, \dots,
$|{\rm A}_{N}\rangle_{p}$, $|{\rm B}_{N}\rangle_{p}$ ($p = e$ or $h$)
is chosen, $\bar{H}_{+}$ is expressed as~\cite{koshino}
\begin{align}
      \label{eq:H_pm}
 \bar{H}_{+}
         = \left(
             \begin{array}{ccccc}
               H_{1} & V & 0_{2\times 2} & 0_{2\times 2} & \cdots \\
               V^{\dagger} & H_{1} & V^{\dagger} & 0_{2\times 2} & \cdots \\
               0_{2\times 2} & V & H_{1} & V & \cdots \\
               0_{2\times 2} & 0_{2\times 2} & V^{\dagger} & H_{1} & \cdots \\
               \cdots & \cdots & \cdots & \cdots & \cdots 
             \end{array}
           \right) ,
\end{align}
where $H_{1}$ is the low-energy effective Hamiltonian
for a monolayer graphene sheet~\cite{slonczewski}
and $V$ describes the inter-layer coupling.
Here $H_{1}$ and $V$ are
\begin{align}
          \label{eq:H1}
  H_{1}   & = \left( \begin{array}{cc}
                       -\tilde{\mu} & \gamma \hat{k}_{-} \\
                       \gamma \hat{k}_{+} & -\tilde{\mu}
                     \end{array}
              \right) ,
          \\
     V    & = \left( \begin{array}{cc}
                        0 & \gamma_{1} \\
                        0 & 0
                     \end{array}
              \right) ,
\end{align}
where $\gamma = (\sqrt{3}/2)\gamma_{0}a$ and
$\hat{k}_{\pm} = \hat{k}_{x} \pm{\rm i}\hat{k}_{y}$ with
$\hat{k}_{x} = -{\rm i}\partial_{x}$ and $\hat{k}_{y} = -{\rm i}\partial_{y}$.
The matrix representation of $\bar{H}_{-}$ is simply obtained by
replacing $\hat{k}_{\pm} \to \hat{k}_{\mp}$ in $\bar{H}_{+}$.

In the presence of superconducting proximity effect,
one must treat the electron and hole states on the same footing,
taking their coupling into account.
For that we focus on the electron-hole space spanned by $|\Psi_{+}\rangle_{e}$
and $|\Psi_{+}\rangle_{h}$.
A natural, and certainly a possible way to proceed is
to employ a Bogoliubov-de Gennes equation in the basis set
adapted for the $N$-layer SGS junction,
$|{\rm A}_{1}\rangle_{e}$, $|{\rm B}_{1}\rangle_{e}$, \dots,
$|{\rm A}_{N}\rangle_{e}$, $|{\rm B}_{N}\rangle_{e}$,
$|{\rm A}_{1}\rangle_{h}$, $|{\rm B}_{1}\rangle_{h}$, \dots,
$|{\rm A}_{N}\rangle_{h}$, $|{\rm B}_{N}\rangle_{h}$.
This has been indeed a conventional approach adopted for this system 
by several authors.~\cite{titov}
To be explicit, the Bogoliubov-de Gennes Hamiltonian in this conventional
approach reads,
\begin{align}
 \underline{H}^{\rm BdG}
   = \left( \begin{array}{cc}
              \bar{H}_{+}
              & \Delta_{\rm eff}(x)\bar{\delta}_{2N\times2N} \\
              \Delta_{\rm eff}(x)^{*}\bar{\delta}_{2N\times2N}
              & -\bar{H}_{+}
            \end{array}
     \right) ,
\end{align}
where $\bar{\delta}_{2N\times2N} = {\rm diag}(1,1,0,\dots,0)$.
Here $\Delta_{\rm eff}(x)$ is the effective pair potential,
and $\bar{\delta}_{2N\times2N}$ represents the fact that only the first
(top) layer is directly coupled with the superconductors.~\cite{takane2}
This conventional approach, however, has a drawback that
it is not appropriate for studying the $T$- and $\Gamma$-dependences
of the Josephson effect taking a proper account of
the planar structure of the junction.
This point has already been mentioned in \S1.

Here, instead of naively applying the Bogoliubov-de Gennes equation
we employ the tunneling Hamiltonian model proposed by McMillan~\cite{mcmillan}.
This approach is especially suited for a proper description of 
the coupling of the graphene sheet and
the superconductors.
The central theoretical tool of this approach is
the thermal Green's function
$\underline{G}(\mib{r},\mib{r}';\omega)$
with the Matsubara frequency $\omega = (2n+1)\pi T$.
In this framework, the proximity effect is mediated by quasiparticle tunneling
between the graphene sheet and the superconductors,
and is described by a self-energy.
The thermal Green's function obeys
\begin{align}
       \label{eq:Green-N}
  \left( {\rm i}\omega\underline{\tau}^{z}-\underline{H}-\underline{\Sigma}
  \right) \underline{G}(\mib{r},\mib{r}';\omega)
  = \underline{\tau}^{0}\delta(\mib{r}-\mib{r}') ,
\end{align}
where $\underline{\tau}^{0}
= {\rm diag}(\bar{1}_{2N \times 2N},\bar{1}_{2N \times 2N})$ and
$\underline{\tau}^{z} = (\bar{1}_{2N \times 2N},-\bar{1}_{2N \times 2N})$
with $\bar{1}_{2N \times 2N}$ being the $2N \times 2N$ unit matrix,
and $\underline{H} = {\rm diag}(\bar{H}_{+}, \bar{H}_{+})$.
The self-energy $\underline{\Sigma}$ is represented as~\cite{takane1}
\begin{align}
 \underline{\Sigma}
           & = \frac{-{\rm i}\Gamma}{\sqrt{\Delta^{2}+\omega^{2}}}
               \left( \begin{array}{cc}
                        \omega\bar{\delta}_{2N\times2N}
                        & \Delta(x)\bar{\delta}_{2N\times2N} \\
                        \Delta(x)^{*}\bar{\delta}_{2N\times2N}
                        & -\omega\bar{\delta}_{2N\times2N}
                      \end{array}
               \right)
                 \nonumber \\
           & \hspace{20mm} \times
               \theta\left(|x|-\frac{L}{2}\right) ,
\end{align}
where $\Gamma$ represents the strength of the tunnel coupling,
and $\theta(x)$ is Heaviside step function.
The off-diagonal elements are regarded as an energy-dependent
effective pair potential,
while the diagonal elements describe renormalization of a quasiparticle energy.
If the $\omega$-dependence is ignored in the self-energy
by setting $\omega = 0$, our model is reduced to the conventional one
with the effective pair potential $\Gamma$.~\cite{titov}

\section{Monolayer and Bilayer Cases}

We derive a general formula for the Josephson current
in the monolayer and bilayer cases
by using an analytical expression of the thermal Green's function.
We employ the quasiclassical approach~\cite{eilenberger,larkin}
and focus on the slowly varying part of the Green's function,
discarding the fast oscillating components on the order of
the Fermi wave number.
This allows for obtaining the Green's function analytically
retaining sufficient accuracy
at distances much longer than the Fermi wave length.

\subsection{Quasiclassical approximation}

By its construction, the Green's function
$\underline{G}(\mib{r},\mib{r}';\omega)$
inherits the matrix nature of the tight-binding (effective mass) Hamiltonian.
In the case of monolayer graphene, it takes a $4\times4$ matrix form
reflecting the sublattice and the electron-hole degrees of freedom.
In the case of bilayer, the size of the matrix is further doubled ($8\times8$)
by the layer degree of freedom.
The Green's function $\underline{G}(\mib{r},\mib{r}';\omega)$
in the bilayer case is defined in the electron-hole space spanned by
$|{\rm A}_{1}\rangle_{e}$, $|{\rm B}_{1}\rangle_{e}$,
$|{\rm A}_{2}\rangle_{e}$, $|{\rm B}_{2}\rangle_{e}$,
$|{\rm A}_{1}\rangle_{h}$, $|{\rm B}_{1}\rangle_{h}$,
$|{\rm A}_{2}\rangle_{h}$, $|{\rm B}_{2}\rangle_{h}$.
However, within the low-energy regime of $|\epsilon| \ll \gamma_{1}$,
we apply a second order perturbation theory~\cite{maccann}
to the $8\times8$ Green's function
and reduce it to a $4\times4$ function
defined in the reduced electron-hole space spanned by
$|{\rm B}_{1}\rangle_{e}$, $|{\rm A}_{2}\rangle_{e}$,
$|{\rm B}_{1}\rangle_{h}$, $|{\rm A}_{2}\rangle_{h}$.
This allows us to treat the mono- and bi-layer cases in parallel;
the Green's function in the two cases are both represented by 
a $4 \times 4$ matrix form,
which we denote as $\check{G}_{j}(\mib{r},\mib{r}';\omega)$.
The subscript $j$ represent the number of layers: $j =1,2$.
The Green's function $\check{G}_{j}(\mib{r},\mib{r}';\omega)$ obeys
\begin{align}
      \label{eq:eq-G_j}
  \left( {\rm i}\omega \check{\tau}_{4\times4}^{z}-\check{H}_{j}
          -\check{\Sigma}_{j}
  \right) \check{G}_{j}(\mib{r},\mib{r}';\omega)
  = \check{\tau}_{4\times4}^{0}\delta(\mib{r}-\mib{r}'),
\end{align}
where $\check{\tau}_{4\times4}^{z} = {\rm diag}(1,1,-1,-1)$,
$\check{\tau}_{4\times4}^{0}={\rm diag}(1,1,1,1)$, and
$\check{H}_{j} = {\rm diag}(H_{j}, H_{j})$.
The $2 \times 2$ effective Hamiltonian $H_{j}$
is given in eq.~(\ref{eq:H1}) in the case of monolayer,
whereas in the case of bilayer
it takes the following form:~\cite{maccann}
\begin{align}
      \label{eq:H2}
  H_{2}
  = \left( \begin{array}{cc}
             -\tilde{\mu} & - \alpha \hat{k}_{+}^{2} \\
             - \alpha \hat{k}_{-}^{2} & -\tilde{\mu}
           \end{array}
    \right) ,
\end{align}
where $\alpha = \gamma^{2}/\gamma_{1}$.
The self-energy $\check{\Sigma}_{j}$ is given by~\cite{takane1}
\begin{align}
     \label{eq:self-e}
 \check{\Sigma}_{j}
     & = \frac{-{\rm i}\Gamma}{\sqrt{\Delta^{2}+\omega^{2}}}
         \left( \begin{array}{cc}
                   \omega \delta_{2 \times 2}^{(j)}
                     & \Delta(x) \delta_{2 \times 2}^{(j)} \\
                   \Delta(x)^{*} \delta_{2 \times 2}^{(j)}
                     & -\omega \delta_{2 \times 2}^{(j)}
                \end{array}
         \right)
                 \nonumber \\
     &   \hspace{20mm} \times
         \theta\left(|x|-\frac{L}{2}\right) ,
\end{align}
where $\delta_{2 \times 2}^{(1)} = {\rm diag}(1,1)$,
and $\delta_{2 \times 2}^{(2)} = {\rm diag}(1,0)$.
The matrix form of $\delta_{2 \times 2}^{(2)}$ reflects the fact that
only the top layer is in contact with superconductors in the bilayer case.

Hereafter, we restrict our attention to the regime of moderate doping:
$\gamma_{0}, \gamma_{1} \gg \mu+U > \mu \gg \Delta$.
Assuming that our system is translationally invariant in the $y$ direction
under the condition of $W \gg L$,
we perform the Fourier transformation:
\begin{align}
  \check{G}_{j}(x,x';q,\omega)
 = \int {\rm d}(y-y') {\rm e}^{-{\rm i}q(y-y')}
   \check{G}_{j}(\mib{r},\mib{r}';\omega) .
\end{align}
The fast spatial oscillations of $\check{G}_{j}$ in the $x$ direction
are characterized by the wave number $k$, which is given by
$k = \sqrt{(\tilde{\mu}/\gamma)^{2}-q^{2}}$ for the monolayer,
and by
$k = \sqrt{\tilde{\mu}/\alpha-q^{2}}$ for the bilayer cases.
Note that this value differs in the covered and uncovered regions
due to the spatial dependence of $\tilde{\mu}$.
For later convenience we introduce the phase
$\chi_{q} \equiv {\rm arg}\{ k+{\rm i}q\}$
in terms of which $k$ and $q$ are expressed as $k = k_{\rm F}\cos\chi_{q}$
and $q = k_{\rm F}\sin\chi_{q}$, respectively,
where $k_{\rm F} = \tilde{\mu}/\gamma$ for the monolayer case
and $k_{\rm F} = \sqrt{\tilde{\mu}/\alpha}$ for the bilayer case.
We separate fast oscillations of $\check{G}_{j}$ by expressing it
as~\cite{zaitsev}
\begin{align}
  \check{G}_{j}(x,x';q,\omega)
 = \sum_{\sigma,\sigma'=\pm}
   \check{G}_{j\sigma\sigma'}(x,x';q,\omega)
   {\rm e}^{{\rm i}k(\sigma x-\sigma' x')} .
\end{align}
Within the accuracy of the quasiclassical approximation,
the Green's function
$\check{G}_{j\sigma,\sigma'}$ obeys
\begin{align}
&  \left( {\rm i}\omega \check{\tau}_{4\times4}^{z}-\check{\mathcal{H}}_{j}
           -\check{\Sigma}_{j}
   \right) \check{G}_{j\sigma\sigma'}(x,x';q,\omega)
                 \nonumber \\
& \hspace{30mm}
   = \delta_{\sigma,\sigma'}\delta(x-x')\check{\tau}_{4\times4}^{0} ,
\end{align}
where the $4 \times 4$ Hamiltonian $\check{\mathcal{H}}_{j}$ is given by
$\check{\mathcal{H}}_{j} = {\rm diag} ( \tilde{H}_{j}+h_{j},
\tilde{H}_{j}+h_{j} )$ with
\begin{align}
 \tilde{H}_{1}
     & = \left( \begin{array}{cc}
                  -\tilde{\mu} & \gamma (\sigma k -{\rm i}q) \\
                   \gamma (\sigma k +{\rm i}q) & -\tilde{\mu}
                \end{array}
         \right) ,
         \\
 h_{1}
     & = \left( \begin{array}{cc}
                   0 & \gamma \hat{k}_{x} \\
                   \gamma \hat{k}_{x} & 0
                \end{array}
         \right) ,
\end{align}
and
\begin{align}
 \tilde{H}_{2}
     & = \left( \begin{array}{cc}
                  -\tilde{\mu} & -\alpha (\sigma k+{\rm i}q)^{2} \\
                  -\alpha (\sigma k-{\rm i}q)^{2} & -\tilde{\mu}
                \end{array}
         \right) ,
         \\
 h_{2}
     & = \left( \begin{array}{cc}
                   0 & -2\alpha (\sigma k +{\rm i}q) \hat{k}_{x} \\
                   -2\alpha (\sigma k -{\rm i}q) \hat{k}_{x} & 0
                \end{array}
         \right) .
\end{align}
Although both the conduction and valence bands contribute to
$\check{G}_{j\sigma\sigma'}$, the latter is irrelevant in considering
low-energy properties since $\tilde{\mu} \gg \Delta$.
Therefore we exclude the contribution from the valence band
by using a unitary transformation which diagonalizes $\tilde{H}_{j}$.
A pair of the eigenvectors, $\psi_{j\sigma}^{\rm c}(q)$
and $\psi_{j\sigma}^{\rm v}(q)$, of $\tilde{H}_{j}$ are given by
\begin{align}
         \label{eq:psi_1}
   \psi_{1\sigma}^{\rm c}(q)
 & = \frac{1}{\sqrt{2}}
     \left( \begin{array}{cc}
              {\rm e}^{-{\rm i}\sigma \chi_{q}/2} \\
              \sigma {\rm e}^{{\rm i}\sigma \chi_{q}/2}
            \end{array}
     \right) ,
    \nonumber \\
   \psi_{1\sigma}^{\rm v}(q)
 & = \frac{1}{\sqrt{2}}
     \left( \begin{array}{cc}
               {\rm e}^{-{\rm i}\sigma \chi_{q}/2} \\
               -\sigma {\rm e}^{{\rm i}\sigma \chi_{q}/2}
            \end{array}
     \right) ,
         \\
         \label{eq:psi_2}
   \psi_{2\sigma}^{\rm c}(q)
 & = \frac{1}{\sqrt{2}}
     \left( \begin{array}{cc}
              {\rm e}^{{\rm i}\sigma \chi_{q}} \\
               -{\rm e}^{-{\rm i}\sigma \chi_{q}}
            \end{array}
     \right) ,
    \nonumber \\
   \psi_{2\sigma}^{\rm v}(q)
 & = \frac{1}{\sqrt{2}}
     \left( \begin{array}{cc}
               {\rm e}^{{\rm i}\sigma \chi_{q}} \\
               {\rm e}^{-{\rm i}\sigma \chi_{q}}
            \end{array}
     \right) ,
\end{align}
and their eigenvalues are $0$ and $-2\tilde{\mu}$, respectively.
Hence $\tilde{H}_{j}$ is diagonalized as
$u_{j\sigma}(q)^{\dagger} \tilde{H}_{j} u_{j\sigma}(q)
= {\rm diag} \left( 0, -2\tilde{\mu} \right)$ with
$u_{j\sigma}(q) = (\psi_{j\sigma}^{\rm c}(q), \psi_{j\sigma}^{\rm v}(q))$,
where the eigenvalue $-2\tilde{\mu}$ corresponds to the valence band.
To exclude the irrelevant contribution from the valence band,
we perform the transformation
$\check{\mathcal{G}}_{j\sigma\sigma'} = \check{U}_{j\sigma}^{\dagger}(q)
\check{G}_{j\sigma\sigma'} \check{U}_{j\sigma'}(q)$
with $\check{U}_{j\sigma}(q) = {\rm diag}(u_{j\sigma}(q), u_{j\sigma}(q))$
and retain only its $(1,1)$-, $(1,3)$-, $(3,1)$-,
and $(3,3)$-elements.~\cite{takane1,takane2}
Accordingly, we define $G_{j\sigma\sigma'}$ by
\begin{align}
 G_{j\sigma\sigma'}
 = \left( \begin{array}{cc}
             \bigl[\check{\mathcal{G}}_{j\sigma\sigma'}\bigr]_{1,1}
             & \bigl[\check{\mathcal{G}}_{j\sigma\sigma'}\bigr]_{1,3} \\
             \bigl[\check{\mathcal{G}}_{j\sigma\sigma'}\bigr]_{3,1}
             & \bigl[\check{\mathcal{G}}_{j\sigma\sigma'}\bigr]_{3,3}
          \end{array}
   \right)
\end{align}
which approximately satisfies the equation of motion
for the quasiclassical Green's function~\cite{eilenberger,larkin}
\begin{align}
      \label{eq:qc-G}
& \left[ {\rm i}\omega \tau_{2\times2}^{z}
         + {\rm i}\sigma v_{jx} \partial_{x}\tau_{2\times2}^{0}
         - \Sigma_{j}
  \right] G_{j\sigma\sigma'}(x,x';q,\omega)
         \nonumber \\
& \hspace{30mm}
  = \delta_{\sigma,\sigma'}\delta(x-x')\tau_{2\times2}^{0} ,
\end{align}
where $\tau_{2\times2}^{z} = {\rm diag}(1,-1)$,
$\tau_{2\times2}^{0} = {\rm diag}(1,1)$, and $v_{jx}$ is the $x$-component
of the velocity given by $v_{jx} = v_{j{\rm F}} \cos\chi_{q}$
with $v_{1{\rm F}} = \gamma$ and $v_{2{\rm F}} = 2\sqrt{\alpha \tilde{\mu}}$.
The self-energy is
\begin{align}
 \Sigma_{j}
   = \frac{-{\rm i}\Gamma_{j}}{\sqrt{\Delta^{2}+\omega^{2}}}
     \left( \begin{array}{cc}
               \omega & \Delta(x) \\
               \Delta(x)^{*} & -\omega
            \end{array}
     \right)
     \theta\left(|x|-\frac{L}{2}\right) ,
\end{align}
where $\Gamma_{1} = \Gamma$ and $\Gamma_{2} = \Gamma/2$.
Note that $\Sigma_{2}$ is smaller by a factor of two than $\Sigma_{1}$
reflecting the fact that only the top layer is in contact
with the superconductors.

Within the approximations described above,
$\check{G}_{j\sigma\sigma'}$ is expressed as $\check{G}_{j\sigma\sigma'}
= G_{j\sigma\sigma'} \otimes \Lambda_{j\sigma\sigma'}$, where
$\Lambda_{j\sigma\sigma'}=u_{j\sigma}(q) \Lambda u_{j\sigma}^{\dagger}(q)$
with $\Lambda = {\rm diag}(1, 0)$.
The explicit forms of $\Lambda_{j\sigma\sigma'}$ are
\begin{align}
& \Lambda_{1\pm\pm}
     = \frac{1}{2}  
       \left( \begin{array}{cc}
                       1 & \pm {\rm e}^{\mp {\rm i}\chi_{q}} \\
                       \pm {\rm e}^{\pm {\rm i}\chi_{q}} & 1
              \end{array}
       \right) ,
      \nonumber \\
& \Lambda_{1\pm\mp}
     = \frac{1}{2}  
       \left( \begin{array}{cc}
                       {\rm e}^{\mp {\rm i}\chi_{q}} & \mp 1 \\
                       \pm 1 & -{\rm e}^{\pm {\rm i}\chi_{q}}
                     \end{array}
              \right)
\end{align}
for the monolayer case, and
\begin{align}
& \Lambda_{2\pm\pm}
     = \frac{1}{2}  
       \left( \begin{array}{cc}
                       1 & -{\rm e}^{\pm {\rm i}2\chi_{q}} \\
                       -{\rm e}^{\mp {\rm i}2\chi_{q}} & 1
              \end{array}
       \right) ,
      \nonumber \\
& \Lambda_{2\pm\mp}
     = \frac{1}{2}  
       \left( \begin{array}{cc}
                       {\rm e}^{\pm {\rm i}2\chi_{q}} & -1 \\
                       -1 & {\rm e}^{\mp {\rm i}2\chi_{q}}
                     \end{array}
              \right)
\end{align}
for the bilayer case.
Consequently, $\check{G}_{j}$ is represented as follows:
\begin{align}
      \label{eq:form-check_G}
& \check{G}_{j}(x,x';q,\omega)
     \nonumber \\
& \hspace{0mm}
 = \sum_{\sigma,\sigma'=\pm} G_{j\sigma\sigma'}(x,x';q,\omega)
   \otimes \Lambda_{j\sigma\sigma'} {\rm e}^{{\rm i}k(\sigma x -\sigma' x')} .
\end{align}
It is convenient to decompose $\check{G}_{j}$ into
four $2 \times 2$ Green's functions as
\begin{align}
  \check{G}_{j}(x,x';q,\omega)
  = \left( \begin{array}{cc}
             g_{j}(x,x';q,\omega) & f_{j}(x,x';q,\omega) \\
             f_{j}^{\dagger}(x,x';q,\omega) & -g_{j}(x,x';q,\omega)
           \end{array}
    \right) .
\end{align}
To obtain a formula for the Josephson current, we need only
$g_{j}$ and $f_{j}^{\dagger}$.
Using eqs.~(\ref{eq:qc-G}) and (\ref{eq:form-check_G})
we can show that in the uncovered region of $-L/2 < x < L/2$,
they are expressed as follows:
\begin{align}
       \label{eq:g-exp}
 & g_{j}(x,x';q,\omega)
      \nonumber \\
 & \hspace{7mm}
   = v_{jx}^{-1}e^{{\rm i}k^{+}(x-x')} \Lambda_{j++}
     \left[ - {\rm i}\theta(x-x') + c_{j++} \right]
      \nonumber \\
 & \hspace{7mm}
    + v_{jx}^{-1}e^{{\rm i}k^{+}(x+x')} \Lambda_{j+-} c_{j+-}
      \nonumber \\
 & \hspace{7mm}
    + v_{jx}^{-1}e^{-{\rm i}k^{+}(x+x')} \Lambda_{j-+} c_{j-+}
      \nonumber \\
 & \hspace{7mm}
    + v_{jx}^{-1}e^{-{\rm i}k^{+}(x-x')} \Lambda_{j--}
      \left[ - {\rm i}\theta(x'-x) + c_{j--} \right] ,
               \\
       \label{eq:f-exp}
 & f_{j}^{\dagger}(x,x';q,\omega)
      \nonumber \\
 & \hspace{7mm}
   =   v_{jx}^{-1}e^{{\rm i}(k^{-}x - k^{+}x')} \Lambda_{j++} d_{j++}
      \nonumber \\
 & \hspace{7mm}
    + v_{jx}^{-1}e^{{\rm i}(k^{-}x + k^{+}x')} \Lambda_{j+-} d_{j+-}
      \nonumber \\
 & \hspace{7mm}
    + v_{jx}^{-1}e^{-{\rm i}(k^{-}x + k^{+}x')} \Lambda_{j-+} d_{j-+}
      \nonumber \\
 & \hspace{7mm}
    + v_{jx}^{-1}e^{-{\rm i}(k^{-}x - k^{+}x')} \Lambda_{j--} d_{j--} ,
\end{align}
where $k^{\pm} = k \pm {\rm i}\omega/v_{jx}$, and $c_{j\sigma\sigma'}$
and $d_{j\sigma\sigma'}$ are unknown coefficients.

\subsection{General formula for the Josephson current}

We show that the Josephson current $I_{j}(\varphi)$ is expressed in terms of
$c_{j++}$ and $c_{j--}$, and then determine these coefficients by applying
a boundary condition at $x = \pm L/2$ to $g_{j}$ and $f_{j}^{\dagger}$.
We finally derive a general formula for the Josephson current.

The Josephson current is formally expressed as
\begin{align}
        \label{eq:I_start_0}
  I_{j}(\varphi)
  = 4W \int\frac{{\rm d}q}{2\pi}T\sum_{\omega}
    {\rm tr}\left\{ \hat{J}_{jx}g_{j}(x;q,\omega)
            \right\} ,
\end{align}
where the factor $4$ comes from the spin and valley degeneracies,
the current operator $\hat{J}_{jx}$ is defined by
\begin{align}
   \hat{J}_{1x}
 & = e\gamma
     \left( \begin{array}{cc}
              0 & 1 \\
              1 & 0
            \end{array}
     \right) ,
               \\
   \hat{J}_{2x}
 & = -2e\alpha
      \left( \begin{array}{cc}
              0 & \hat{k}_{x}+{\rm i}q \\
              \hat{k}_{x}-{\rm i}q & 0
            \end{array}
     \right) ,
\end{align}
and $g_{j}(x;q,\omega) \equiv [g_{j}(x,x-0;q,\omega)+g_{j}(x,x+0;q,\omega)]/2$.
Substituting eq.~(\ref{eq:g-exp}) into eq.~(\ref{eq:I_start_0})
and changing the integration variable from $q$ to $\chi_{q}$, we obtain
\begin{align}
        \label{eq:I_start}
  I_{j}(\varphi)
 & = 2eN_{j}(0)W
      \nonumber \\
 & \hspace{2mm} \times
     \int_{-\frac{\pi}{2}}^{+\frac{\pi}{2}}
     {\rm d}\chi_{q} v_{jx}
     T\sum_{\omega} \left(c_{j++}(\varphi)-c_{j--}(\varphi)\right) ,
\end{align}
where $N_{j}(0)$ the density of states per spin
including the valley degeneracy is given by
$N_{1}(0) = \mu/(\pi \gamma^{2})$ and $N_{2}(0) = 1/(2\pi \alpha)$.

The unknown coefficients are determined by the boundary condition
at $x = \pm L/2$ for $g_{j}(x,x';q,\omega)$ and
$f_{j}^{\dagger}(x,x';q,\omega)$,~\cite{galaktionov} which we outline below.
It is necessary to distinguish $\chi_{q}$ in the covered and uncovered regions
in the following argument, so we rewrite $\chi_{q}$
as $\theta$ ($\phi$) in the covered (uncovered) region:
\begin{align}
      \label{eq:def-phi/theta}
  \chi_{q}
  = \left\{ \begin{array}{cc}
               \phi & (-L/2 < x < L/2) \\
               \theta  & (L/2 < |x|)
            \end{array}
    \right. .
\end{align}
To represent the boundary condition we introduce electron wave function
$\Psi_{{\rm N}j}^{\rm e}(x;q,\omega)$ and hole wave function
$\Psi_{{\rm N}j}^{\rm h}(x;q,\omega)$ in the uncovered region.
They are decomposed into right-going and left-going components as
\begin{align}
  \Psi_{{\rm N}j}^{\rm e}(x;q,\omega)
 & = c_{\rm N+}(x,\omega)\psi_{j+}^{\rm c}(q)
     + c_{\rm N-}(x,\omega)\psi_{j-}^{\rm c}(q) ,
        \\
  \Psi_{{\rm N}j}^{\rm h}(x;q,\omega)
 & = d_{\rm N+}(x,\omega)\psi_{j+}^{\rm c}(q)
     + d_{\rm N-}(x,\omega)\psi_{j-}^{\rm c}(q) ,
\end{align}
where $c_{\rm N\pm}(x,\omega)$ ($d_{\rm N\pm}(x,\omega)$) contains
the $x$-dependent factor of ${\rm e}^{\pm{\rm i}k^{+}x}$
(${\rm e}^{\pm{\rm i}k^{-}x}$).
Applying the prescription by Zaitsev,~\cite{zaitsev}
we derive a relation between $c_{\rm N\pm}(\pm L/2,\omega)$
and $d_{\rm N\pm}(\pm L/2,\omega)$.
We leave its derivation to Appendix A, and here
refers only to the final result,
\begin{align}
        \label{eq:BC-condition}
     \left( \begin{array}{c}
               c_{\rm N+}(\pm L/2,\omega) \\
               c_{\rm N-}(\pm L/2,\omega)
            \end{array}
     \right)
   = B_{j}(\pm L/2)
     \left( \begin{array}{c}
               d_{\rm N+}(\pm L/2,\omega) \\
               d_{\rm N-}(\pm L/2,\omega)
            \end{array}
     \right)
\end{align}
with
\begin{align}
  B_{j}(\pm L/2) =
  \frac{{\rm e}^{\pm {\rm i}\varphi/2}}{\tilde{\Delta}}
  \left(
    \begin{array}{cc}
      \tilde{\omega}\pm\frac{1+\mathcal{R}_{j}}{\mathcal{T}_{j}}\Omega
      & \mp\frac{2\sqrt{\mathcal{R}_{j}}{\rm e}^{-{\rm i}\zeta}}
                {\mathcal{T}_{j}}\Omega \\
      \pm\frac{2\sqrt{\mathcal{R}_{j}}{\rm e}^{{\rm i}\zeta}}
              {\mathcal{T}_{j}}\Omega
      & \tilde{\omega}\mp\frac{1+\mathcal{R}_{j}}{\mathcal{T}_{j}}\Omega
    \end{array}
  \right) ,
\end{align}
where
\begin{align}
  \tilde{\omega}
  & = \left(1+\frac{\Gamma_{j}}{\sqrt{\omega^{2}+\Delta^{2}}} \right)\omega ,
          \\
  \tilde{\Delta}
  & = \frac{\Gamma_{j}}{\sqrt{\omega^{2}+\Delta^{2}}} \Delta ,
          \\
  \Omega
  & = \sqrt{\tilde{\omega}^{2}+\tilde{\Delta}^{2}} ,
\end{align}
and $\mathcal{T}_{j}$ and $\mathcal{R}_{j} \equiv 1-\mathcal{T}_{j}$
denote the transmission and reflection probabilities
for an electron at the Fermi level being incident at
the uncovered-to-covered interface with an incident angle $\phi$,
and $\zeta$ is the phase of the corresponding reflection coefficient.
The transmission probability for the monolayer case is
\begin{align}
      \label{eq:T1}
  \mathcal{T}_{1} = \frac{2\cos\phi\cos\theta}{1+\cos(\phi+\theta)} ,
\end{align}
and that for the bilayer case is
\begin{align}
      \label{eq:T2}
  \mathcal{T}_{2}
  = \frac
    {4\left(1+\mathcal{L}\right)^{2}w\cos\phi\cos\theta}
    {\left(1+\mathcal{L}^{2}\right)A + 2\mathcal{L}B}
\end{align}
with
\begin{align}
  A & = 1+w^{2}+2w\cos(\phi+\theta) , \\
  B & = \cos 2\phi+w^{2}\cos 2\theta+2w\cos(\phi-\theta) ,
\end{align}
where $w = \sqrt{\mu/(\mu+U)}$ and
\begin{align}
      \label{eq:def-L}
  \mathcal{L}
  = \frac
    {\left(\sqrt{1+\sin^{2}\phi}+\sin\phi\right)
     \left(\sqrt{1+\sin^{2}\phi+U/\mu}-\sin\phi\right)}
    {\left(\sqrt{1+\sin^{2}\phi}-\sin\phi\right)
     \left(\sqrt{1+\sin^{2}\phi+U/\mu}+\sin\phi\right)} .
\end{align}
The derivation of eq.~(\ref{eq:T1}) and (\ref{eq:T2}) is given in Appendix B.
We determine the unknown coefficients
in eqs.~(\ref{eq:g-exp}) and (\ref{eq:f-exp})
by using eq.~(\ref{eq:BC-condition}) as the boundary condition.
Equation~(\ref{eq:BC-condition}) provides us coupled eight relations,
two of which are
\begin{align}
     \left( \begin{array}{c}
               (-{\rm i}+c_{j++}){\rm e}^{{\rm i}k^{+}L/2} \\
               c_{j-+}{\rm e}^{-{\rm i}k^{+}L/2}
            \end{array}
     \right)
   = B_{j}(L/2)
     \left( \begin{array}{c}
               d_{j++}{\rm e}^{{\rm i}k^{-}L/2} \\
               d_{j-+}{\rm e}^{-{\rm i}k^{-}L/2}
            \end{array}
     \right) .
\end{align}
Solving these coupled relations we obtain
\begin{align}
 c_{j++}(\varphi)
 = \frac{1}{2{\rm i}}\left(\frac{\Upsilon_{j}}
                                {\Pi_{j}}-1\right)
\end{align}
with
\begin{align}
      \label{eq:Upsilon}
 \Upsilon_{j}
 & =  \left(\tilde{\omega}^{2}+\mathcal{A}_{j}^{2}\Omega^{2}\right)
      \sinh\kappa_{j}
    + 2\mathcal{A}_{j}\tilde{\omega}\Omega\cosh\kappa_{j}
      \nonumber \\
 & \hspace{10mm}
    + {\rm i}\mathcal{B}_{j}^{2}\Omega^{2}\sin2kL
    + {\rm i}\tilde{\Delta}^{2}\sin\varphi ,
         \\
      \label{eq:Pi}
 \Pi_{j}
 & =  \left(\tilde{\omega}^{2}+\mathcal{A}_{j}^{2}\Omega^{2}\right)
      \cosh\kappa_{j}
    + 2\mathcal{A}_{j}\tilde{\omega}\Omega \sinh\kappa_{j}
      \nonumber \\
 & \hspace{10mm}
    - \mathcal{B}_{j}^{2}\Omega^{2}\cos2kL
    + \tilde{\Delta}^{2}\cos\varphi ,
\end{align}
where $\mathcal{A}_{j} = (1+\mathcal{R}_{j})/\mathcal{T}_{j}$,
$\mathcal{B}_{j} = 2\sqrt{\mathcal{R}_{j}}/\mathcal{T}_{j}$,
and $\kappa_{j} \equiv 2\tilde{\omega}L/v_{jx}$.
The other coefficient $c_{j--}(\varphi)$ satisfies
$c_{j--}(\varphi)=c_{j++}(-\varphi)$, so we do not present it explicitly.

Substituting these results into eq.~(\ref{eq:I_start}),
we finally obtain a general formula for the Josephson current
\begin{align}
        \label{eq:I_final}
    I_{j}(\varphi)
    = e\mathcal{N}_{j} \int_{-\frac{\pi}{2}}^{+\frac{\pi}{2}}
      {\rm d}\phi \cos\phi T\sum_{\omega}
      \frac{\tilde{\Delta}^{2}\sin\varphi}{\Pi_{j}} ,
\end{align}
where $\mathcal{N}_{j} \equiv 2v_{j{\rm F}}N_{j}(0)W$
represents the number of conducting channels.
Using this general formula one can numerically calculate
the Josephson current in the planar junction for arbitrary parameters.
If we take the strong-coupling limit of $\Gamma_{j} \gg \Delta_{0}$
($\Delta_{0}$: the magnitude of the pair potential at $T = 0$),
eq.~(\ref{eq:I_final}) is reduced to the expression for
the Josephson current in an SNS junction (N: normal metal)
with same barriers at the two NS interfaces,
derived by Galaktionov and Zaikin.~\cite{galaktionov}

\section{Extension to the $N$-layer Case}

To study the Josephson effect in a planar junction of $N$-layer graphene,
we first decompose the $4N \times 4N$ thermal Green's function
into a set of $4 \times 4$ Green's functions.
With this decomposition we can easily derive a formula for
the Josephson current by applying the argument presented
in the previous section.

Following Koshino and Ando~\cite{koshino}, we decompose
$\bar{H}_{+}$ for an $N$-layer graphene sheet 
into $2 \times 2$ monolayer-type
and $4 \times 4$ bilayer-type Hamiltonians
with the help of an appropriate choice of the bases.
For $N$ being odd, the $2N \times 2N$ Hamiltonian $\bar{H}_{+}$ is
decomposed into one monolayer-type Hamiltonian and 
$(N-1)/2$ bilayer-type Hamiltonians.
While, for $N$ being even,
$\bar{H}_{+}$ is decomposed into $N/2$ bilayer-type Hamiltonians.
This procedure can be adapted to 
a multilayer graphene sheet in {\it planar} contact
with superconductors as demonstrated in ref.~\citen{takane2}.

Applying this decomposition also to the thermal Green's function
$\underline{G}(\mib{r},\mib{r}';\omega)$,
we can decompose it into $4 \times 4$ monolayer-type
and $8 \times 8$ bilayer-type Green's functions.
The obtained $8 \times 8$ bilayer-type Green's function is then 
reduced to a $4 \times 4$ Green's function.
These decomposition and reduction procedures are explained in Appendix C.
In the case of $N$ being odd, 
$\underline{G}(\mib{r},\mib{r}';\omega)$ is decomposed
into one monolayer-type Green's function
$\check{G}_{N,0}(\mib{r},\mib{r}';\omega)$ and $(N-1)/2$ bilayer-type Green's
functions $\check{G}_{N,m}(\mib{r},\mib{r}';\omega)$ with $m = 2,4,\dots, N-1$.
While, for $N$ being even,
$\underline{G}(\mib{r},\mib{r}';\omega)$
is decomposed into $N/2$ bilayer-type Green's functions
$\check{G}_{N,m}(\mib{r},\mib{r}';\omega)$ with $m = 1,3,\dots, N-1$.

The Green's function $\check{G}_{N,m}$ obeys
\begin{align}
      \label{eq:eq-G}
  \left( {\rm i}\omega \check{\tau}_{4\times4}^{z}-\check{H}_{N,m}
          -\check{\Sigma}_{N,m}
  \right) \check{G}_{N,m}(\mib{r},\mib{r}';\omega)
  = \check{\tau}_{4\times4}^{0}\delta(\mib{r}-\mib{r}') .
\end{align}
For the Green's function 
of monolayer-type with $m=0$,
the Hamiltonian and the self-energy are given by
$\check{H}_{N,0} = {\rm diag}(H_{1}, H_{1})$ and
\begin{align}
 \check{\Sigma}_{N,0}
    & = \frac{-{\rm i}\Gamma_{N,0}}{\sqrt{\Delta^{2}+\omega^{2}}}
        \left( \begin{array}{cc}
                  \omega \delta_{2 \times 2}^{(1)}
                  & \Delta(x) \delta_{2 \times 2}^{(1)} \\
                  \Delta(x)^{*} \delta_{2 \times 2}^{(1)}
                  & -\omega \delta_{2 \times 2}^{(1)}
               \end{array}
        \right)
        \nonumber \\
    & \hspace{20mm} \times
        \theta\left(|x|-\frac{L}{2}\right)
\end{align}
with
\begin{align}
  \Gamma_{N,0} = \frac{2}{N+1}\Gamma .
\end{align}
For the bilayer-type with $m \neq 0$,
the Hamiltonian is given by
$\check{H}_{N,m} = {\rm diag}(H_{2}^{N,m}, H_{2}^{N,m})$, where
\begin{align}
      \label{eq:H_2-Nm}
  H_{2}^{N,m}
  = \left( \begin{array}{cc}
             -\tilde{\mu} & - \frac{\alpha}{\lambda_{N,m}} \hat{k}_{+}^{2} \\
             - \frac{\alpha}{\lambda_{N,m}} \hat{k}_{-}^{2} & -\tilde{\mu}
           \end{array}
    \right)
\end{align}
with
\begin{align}
      \label{eq:def-lambda}
 \lambda_{N,m} = 2\sin\left(\frac{m\pi}{2(N+1)}\right) .
\end{align}
The self-energy is given by
\begin{align}
 \check{\Sigma}_{N,m}
    & = \frac{-2{\rm i}\Gamma_{N,m}}{\sqrt{\Delta^{2}+\omega^{2}}}
        \left( \begin{array}{cc}
                  \omega \delta_{2 \times 2}^{(2)}
                    & \Delta(x) \delta_{2 \times 2}^{(2)} \\
                  \Delta(x)^{*} \delta_{2 \times 2}^{(2)}
                    & -\omega \delta_{2 \times 2}^{(2)}
               \end{array}
        \right)
        \nonumber \\
    & \hspace{20mm} \times
        \theta\left(|x|-\frac{L}{2}\right)
\end{align}
with
\begin{align}
  \Gamma_{N,m}
  = \frac{2\cos^{2}\left(\frac{m\pi}{2(N+1)}\right)}{N+1} \Gamma .
\end{align}
The Green's function $\check{G}_{N,m}$ is reduced to
that in the monolayer (bilayer) case if $N = 1$ ($N = 2$).

Applying the procedure described in the previous section to $\check{G}_{N,m}$,
we obtain the corresponding contribution $I_{N,m}(\varphi)$
to the Josephson current, which is given by eq.~(\ref{eq:I_final})
with the following replacements:
\begin{align}
      \label{eq:replace1}
  & v_{j{\rm F}} \to v_{\rm F}^{N,m} ,
           \\
      \label{eq:replace2}
  & N_{j}(0) \to N_{N,m}(0) ,
           \\
      \label{eq:replace3}
  & \Gamma_{j} \to \Gamma_{N,m} ,
\end{align}
where
\begin{align}
  v_{\rm F}^{N,0} & = v_{1\rm F} = \gamma ,
           \\
  v_{\rm F}^{N,m} & = \frac{v_{2\rm F}}{\sqrt{\lambda_{N,m}}}
                    = 2\sqrt{\frac{\alpha\mu}{\lambda_{N,m}}} ,
\end{align}
and
\begin{align}
  N_{N,0}(0) & = N_{1}(0) = \frac{\mu}{\pi \gamma^{2}} ,
           \\
  N_{N,m}(0) & = \lambda_{N,m} N_{2}(0)
               = \frac{\lambda_{N,m}}{2\pi\alpha} .
\end{align}
The total Josephson current $I_{N}(\varphi)$
in the $N$-layer case is expressed as
\begin{align}
     \label{eq:IN_sum}
  I_{N}(\varphi) = \sum_{m}I_{N,m}(\varphi) .
\end{align}

\section{Analytical and Numerical Results}

Let us focus on the short-junction limit: $L \ll \xi$,
where $\xi \equiv v_{j{\rm F}}/(2\pi \Delta_{0})$
is the superconducting coherence length.
This limit has a particular importance to the interpretation of
experiments.
Taking this limit helps simplifying the denominator of eq.~(\ref{eq:I_final}) by
setting $\cosh\kappa_{j} \approx 1$ and $\sinh\kappa_{j} \approx 0$,
resulting in
\begin{align}
        \label{eq:I_short}
    I_{j}(\varphi)
  & = e\mathcal{N}_{j} \int_{-\frac{\pi}{2}}^{+\frac{\pi}{2}}
      {\rm d}\phi \cos\phi T\sum_{\omega}
      \nonumber \\
  & \hspace{-10mm} \times
      \frac{\tilde{\Delta}^{2}\sin\varphi}
      {\tilde{\omega}^{2}
       + \left(\frac{1+\mathcal{R}_{j}}{\mathcal{T}_{j}}\right)^{2}\Omega^{2}
       - \left(\frac{2\sqrt{\mathcal{R}_{j}}}{\mathcal{T}_{j}}\right)^{2}
         \Omega^{2}\cos2kL
       + \tilde{\Delta}^{2}\cos\varphi} .
\end{align}
Based on this formula, we first study analytically the behavior of
the Josephson current in several limiting cases,
then estimate the expression numerically
to observe those features that are not captured by the analytical arguments.

\subsection{Analytical result}

\subsubsection{Cases of homogeneous carrier density}
Let us first consider the simplest case of $U = 0$
(without carrier inhomogeneity).
Since $\mathcal{T}_{j}=1$ and $\mathcal{R}_{j}=0$ in this case,
eq.~(\ref{eq:I_short}) is further simplified to
\begin{align}
        \label{eq:Ij_simplest}
    I_{j}(\varphi)
    = 2e\mathcal{N}_{j}
      T\sum_{\omega}
      \frac{\tilde{\Delta}^{2}\sin\varphi}
           {\tilde{\omega}^{2} + \Omega^{2}+\tilde{\Delta}^{2}\cos\varphi}
\end{align}
for the monolayer and bilayer cases.~\cite{takane1}
The behavior of $I_{j}(\varphi)$ predicted from eq.~(\ref{eq:Ij_simplest})
has been discussed in ref.~\citen{takane1}.
In the strong coupling limit of $\Gamma_{j} \gg \Delta_{0}$, this formula
reproduces the result derived by Kulik and Omel'yanchek (KO),~\cite{kulik}
\begin{align}
       \label{eq:I-nm-KO2}
 I_{j}(\varphi)
   = e\mathcal{N}_{j}\Delta
     \sin\frac{\varphi}{2}
     \tanh\left(\frac{\Delta \cos\frac{\varphi}{2}}{2T}\right) .
\end{align}
In the weak coupling limit of $\Delta_{0} \gg \Gamma_{j}$,
the Josephson current behaves as
\begin{align}
       \label{eq:I-nm-volkov}
 I_{j}(\varphi)
   = e\mathcal{N}_{j}\Gamma_{j}
     \sin\frac{\varphi}{2}
     \tanh\left(\frac{\Gamma_{j} \cos\frac{\varphi}{2}}{2T}\right)
\end{align}
in the low-temperature regime of $\Gamma_{j} \gg T$,~\cite{volkov} and
\begin{align}
       \label{eq:I-nm-Tc}
 I_{j}(\varphi)
   = e\mathcal{N}_{j}
     \frac{\Gamma_{j}^{2}\Delta^{2}}{48T^{3}}\sin\varphi
\end{align}
in the high temperature regime of $T_{\rm c} \gtrsim T$.
Note that the $T$-dependence of $I_{j}(\varphi)$ in the weak coupling limit
is qualitatively differ from the KO result which yields
$I(\varphi)=e\mathcal{N}\Delta^{2}(4T)^{-1}\sin\varphi$ near $T_{\rm c}$.
The behavior revealed in eq.~(\ref{eq:I-nm-Tc}) has not been reported
in literature, indicating that the Josephson effect in a planar junction
is significantly affected by the coupling strength.

\subsubsection{Back to the generics --- cases of inhomogeneous carrier density}

Let us go back to eq.~(\ref{eq:I_short}), and investigate the behavior
of $I_{j}(\varphi)$ in several specific limits.
Rewriting $\tilde{\Delta}^{2}/\Pi_{j}$ in eq.~(\ref{eq:I_short}) as
\begin{align}
    \frac{\tilde{\Delta}^{2}}
         {\Pi_{j}}
  = \frac{(\Gamma_{j}\Delta)^{2}}
         {\Phi_{j}\mathcal{D}_{j}
          -2(\Gamma_{j}\Delta)^{2}\sin^{2}\frac{\varphi}{2}}
\end{align}
with
\begin{align}
  \Phi_{j}
  & = 1 + \left(\frac{1+\mathcal{R}_{j}}{\mathcal{T}_{j}}\right)^{2}
        - \left(\frac{2\sqrt{\mathcal{R}_{j}}}
                            {\mathcal{T}_{j}}\right)^{2}\cos2kL , \\
  \mathcal{D}_{j}
  & = \omega^{4} + \left(\Delta^{2}+2\Gamma_{j}\sqrt{\Delta^{2}+\omega^{2}}
                 + \Gamma_{j}^{2}\right)\omega^{2}
                 + (\Gamma_{j}\Delta)^{2} ,
\end{align}
we approximate $2\Gamma_{j}\sqrt{\Delta^{2}+\omega^{2}}$ in $\mathcal{D}_{j}$
by $2\Gamma_{j}\Delta$.
This is justified in three situations:
the low-temperature limit of $T_{\rm c} \gg T$,
the strong-coupling limit of $\Gamma_{j} \gg \Delta_{0}$,
and the weak-coupling limit of $\Delta_{0} \gg \Gamma_{j}$.
Under this approximation we perform the summation over $\omega$
in eq.~(\ref{eq:I_short}) in terms of a contour integral, and obtain
\begin{align}
        \label{eq:I-approx}
    I_{j}(\varphi)
  & = e\mathcal{N}_{j} (\Gamma_{j}\Delta)^{2}\sin\varphi
      \int_{-\frac{\pi}{2}}^{+\frac{\pi}{2}}
      {\rm d}\phi
      \frac{\cos\phi}{\Phi_{j}E_{j+}E_{j-}}
      \nonumber \\
  & \times
      \left\{  \frac{\tanh\left[\frac{E_{j+}-E_{j-}}{4T}
                          \right]}{E_{j+}-E_{j-}}
             - \frac{\tanh\left[\frac{E_{j+}+E_{j-}}{4T}
                          \right]}{E_{j+}+E_{j-}}
      \right\} ,
\end{align}
where
\begin{align}
  E_{j\pm}
   = \sqrt{(\Delta+\Gamma_{j})^{2}\pm2\sqrt{|f_{j}|}\Gamma_{j}\Delta}
\end{align}
with
\begin{align}
  f_{j}
  = 1 - \frac{2\sin^{2}\frac{\varphi}{2}}{\Phi_{j}} .
\end{align}
Equation~(\ref{eq:I-approx}) reduces to eq.~(31) of ref.~\citen{takane1}
in the homogeneous case of $U = 0$.~\cite{comment2}

The characteristic behavior of $I_{j}(\varphi)$ in the strong- and
weak-coupling limits follows from eq.~(\ref{eq:I-approx}).
The Josephson current in the strong-coupling limit is given by
\begin{align}
       \label{eq:I_app1}
    I_{j}(\varphi)
  = e\mathcal{N}_{j} \Delta\sin\varphi
    \int_{-\frac{\pi}{2}}^{+\frac{\pi}{2}}{\rm d}\phi
    \frac{\cos\phi}{2\Phi_{j}\sqrt{|f_{j}|}}
    \tanh\left(\frac{\sqrt{|f_{j}|}\Delta}{2T}\right)
\end{align}
for an arbitrary $T$.
In the weak-coupling limit, we obtain
\begin{align}
       \label{eq:I_app2}
    I_{j}(\varphi)
  = e\mathcal{N}_{j} \Gamma_{j}\sin\varphi
    \int_{-\frac{\pi}{2}}^{+\frac{\pi}{2}}{\rm d}\phi
    \frac{\cos\phi}{2\Phi_{j}\sqrt{|f_{j}|}}
    \tanh\left(\frac{\sqrt{|f_{j}|}\Gamma_{j}}{2T}\right)
\end{align}
in the low-temperature regime of $\Gamma_{j} \gg T$, and
\begin{align}
       \label{eq:I_app3}
    I_{j}(\varphi)
  = e\mathcal{N}_{j} \frac{c_{j}(\Gamma_{j}\Delta)^{2}\sin\varphi}{48T^{3}}
\end{align}
in the high-temperature regime of $T_{\rm c} \gtrsim T$, where
\begin{align}
  c_{j} = \int_{-\frac{\pi}{2}}^{+\frac{\pi}{2}}{\rm d}\phi
          \frac{\cos\phi}{\Phi_{j}} .
\end{align}
Note that eqs.~(\ref{eq:I_app1}), (\ref{eq:I_app2}), and (\ref{eq:I_app3})
reduce to eqs.~(\ref{eq:I-nm-KO2})-(\ref{eq:I-nm-Tc}), respectively,
in the homogeneous case of $U = 0$.

\subsubsection{Zero-temperature behaviors}

Let us consider the critical current $I_{j{\rm c}}$ at $T = 0$,
where eq.~(\ref{eq:I-approx}) simplifies to
\begin{align}
       \label{eq:I-approx-T=0}
    I_{j}(\varphi)
  & = e\mathcal{N}_{j}
      \frac{\Gamma_{j}\Delta_{0}}{\Delta_{0}+\Gamma_{j}}\sin\varphi
      \int_{-\frac{\pi}{2}}^{+\frac{\pi}{2}}
      {\rm d}\phi
      \nonumber \\
  & \times
      \frac{\cos\phi}{2\Phi_{j}\sqrt{|f_{j}|}
                      \sqrt{1+\frac{2\Gamma_{j}\Delta_{0}}
                                    {(\Delta_{0}+\Gamma_{j})^{2}}
                      \sqrt{|f_{j}|}}} .
\end{align}
In the homogeneous case of $U = 0$,
$\Phi_{j} = 2$ since $\mathcal{T}_{j}=1$ irrespective of $\phi$.
In this case $I_{j}(\varphi)$ is maximized at $\varphi = \pi$ (mod $2\pi$).
Thus the critical current at $U = 0$ is determined as $I_{j{\rm c}}^{U=0}
= e\mathcal{N}_{j}\Gamma_{j}\Delta_{0}/(\Delta_{0}+\Gamma_{j})$.
This yields
\begin{align}
  I_{1{\rm c}}^{U=0}
   = e\frac{2\mu W}{\pi \gamma}
     \frac{\Gamma\Delta_{0}}{\Delta_{0}+\Gamma}
\end{align}
for the monolayer, and
\begin{align}
  I_{2{\rm c}}^{U=0}
   = e\frac{2\mu W}{\pi \gamma}\sqrt{\frac{\gamma_{1}}{\mu}}
     \frac{\frac{\Gamma}{2}\Delta_{0}}{\Delta_{0}+\frac{\Gamma}{2}}
\end{align}
for the bilayer cases.
Speaking of the order of magnitude, 
one can notify that
the critical current in the case of bilayer
is greater than that of monolayer
by a factor of
$\sqrt{\gamma_{1}/\mu}$.

\subsubsection{Cases of multi-layers}

The critical current for the cases of multi-layers at $T = 0$
can be discussed on the same footing.
As demonstrated in eq.~(\ref{eq:IN_sum}),
the Josephson current in the case of $N$-layer
\begin{enumerate}
\item 
is given by the summation of $I_{N,m}(\varphi)$ over $m$,
\item
and every contribution is
maximized at $\varphi = \pi$ when $U = 0$.
\end{enumerate}
Combining these two observations, we see that the critical current
$I_{N{\rm c}}^{U=0}$ for the $N$-layer case is simply given by
$\sum_{m}I_{N,m}(\varphi)|_{\varphi \to \pi}$,
where $I_{N,m}(\varphi)$ is obtained by performing
the replacements, eqs.~(\ref{eq:replace1})-(\ref{eq:replace3}),
in eq.~(\ref{eq:I-approx-T=0}).

Listing up the cases of small $N$, one can find
\begin{align}
  I_{3{\rm c}}^{U=0}
   = e\frac{2\mu W}{\pi \gamma}
     \left[
        \frac{\frac{\Gamma}{2}\Delta_{0}}{\Delta_{0}+\frac{\Gamma}{2}}
      + \sqrt{\frac{\sqrt{2}\gamma_{1}}{\mu}}
        \frac{\frac{\Gamma}{4}\Delta_{0}}{\Delta_{0}+\frac{\Gamma}{4}}
     \right]
\end{align}
for the tri-layer, and
\begin{align}
  I_{4{\rm c}}^{U=0}
 & = e\frac{2\mu W}{\pi \gamma}
     \Bigg[
        \sqrt{\frac{\frac{\sqrt{5}-1}{2}\gamma_{1}}{\mu}}
        \frac{\frac{(1+\sqrt{5}^{-1})\Gamma}{4}\Delta_{0}}
             {\Delta_{0}+\frac{(1+\sqrt{5}^{-1})\Gamma}{4}}
     \nonumber \\
 & \hspace{15mm}
      + \sqrt{\frac{\frac{\sqrt{5}+1}{2}\gamma_{1}}{\mu}}
        \frac{\frac{(1-\sqrt{5}^{-1})\Gamma}{4}\Delta_{0}}
             {\Delta_{0}+\frac{(1-\sqrt{5}^{-1})\Gamma}{4}}
     \Bigg]
\end{align}
for the tetra-layer cases.

\subsection{Numerical result}

The use of the analytic formulas obtained so far
is limited in the vicinity of a particular limit we have specified each time
in the space of control parameters, $(U, T)$.
Here, to give an insight on the behavior of the critical current
in the entire range of this parameter space, we discuss some numerical plots
obtained by directly estimating the expressions, such as
eqs.~(\ref{eq:I-approx-T=0}), (\ref{eq:Ij_simplest}) and (\ref{eq:I_short})
(instead of trying to simplify them
by restricting the range of validity). 
In the actual computation,
the following set of parameters is employed:
$\gamma_{0} = 2.8 \ {\rm eV}$, $\gamma_{1} = 0.4 \ {\rm eV}$,
$a = 0.246 \ {\rm nm}$, $\Delta_{0} = 120 \ \mu{\rm eV}$,
$\mu = 32 \ {\rm meV}$, and $L = 200 \ {\rm nm}$.

%%%%%%%%%%%%%%%%%%%%%%%%
\begin{figure}[btp]
\begin{center}
\includegraphics[height=6.0cm]{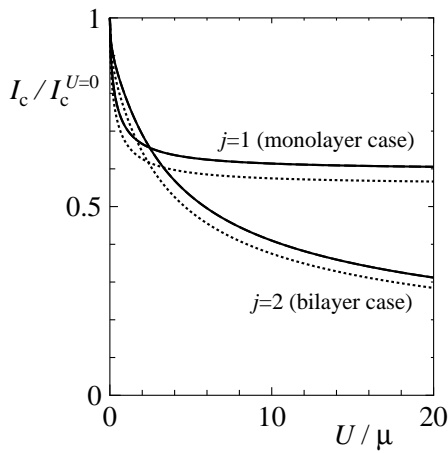}
\end{center}
\caption{The normalized critical current
$I_{j\rm c}/I_{j{\rm c}}^{U=0}$ at $T = 0$
as a function of $U/\mu$, where $r = 0.1$ (dashed lines),
$1.0$ (dotted lines), $10.0$ (solid lines).
The data for $r = 0.1$ and $r = 10.0$ are almost overlapped
and indistinguishable in both the monolayer and bilayer cases.
}
\end{figure}
%%%%%%%%%%%%%%%%%%%%%%

\subsubsection{$U$-dependence of the critical current}

Fixing the temperature at $T = 0$, let us focus on the $U$-dependence of
the critical current in the monolayer and bilayer cases.
At $T=0$ the critical current $I_{j{\rm c}}$ can be computed
at an arbitrary value of $U$, measure of the carrier inhomogeneity,
by numerically estimating eq.~(\ref{eq:I-approx-T=0}).
In Fig.~2 the critical current $I_{j{\rm c}}$ 
(normalized by $I_{j{\rm c}}^{U=0}$) is shown as a function of $U/\mu$
for different values of a parameter $r \equiv \Gamma_{j}/\Delta_{0}$
that characterizes
the coupling strength between the graphene sheet and the superconductors.
In the figure,
different curves, corresponding respectively
to a different choice of this parameter
$r = 0.1$, $1.0$, and $10.0$, and to a different number of layers ($j=1, 2$)
are superposed for comparison.
The behavior of $I_{j{\rm c}}/I_{j{\rm c}}^{U=0}$ shown in the figure
indicates that it is almost insensitive to the variation of $r$ in this range,
but exhibits a qualitatively different behavior
in the monolayer and bilayer cases.

The suppression of $I_{j{\rm c}}/I_{j{\rm c}}^{U=0}$ as the increase of $U$ is
more pronounced in the bilayer case.
In the case of monolayer, the ratio $I_{1{\rm c}}/I_{1{\rm c}}^{U=0}$,
indeed converges to a constant value
in the limit of $U\rightarrow\infty$.~\cite{titov}
On contrary,
$I_{2{\rm c}}/I_{2{\rm c}}^{U=0}$ decreases monotonically toward zero.
This contrasting behavior reflects the difference of
the transmission probability $\mathcal{T}_{j}$ in the two cases.
In the simple case of an electron at the Fermi level
incident perpendicularly to the interface,
the electron is completely transmitted
in monolayer even though $U$ is very large.
Contrastingly, the transmission probability significantly decreases
with increasing $U$ in the bilayer case.

%%%%%%%%%%%%%%%%%%%%%%%
\begin{figure}[btp]
\begin{center}
\includegraphics[height=5.5cm]{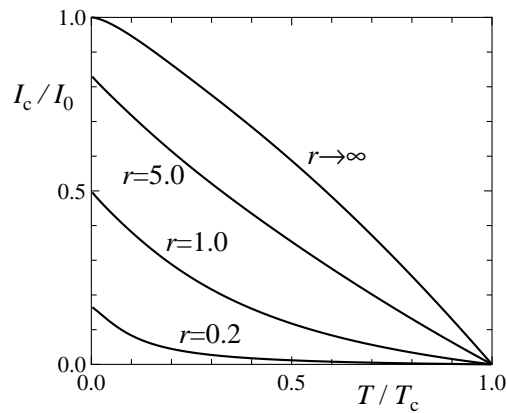}
\end{center}
\caption{The normalized critical current $I_{j{\rm c}}/I_{0}^{(j)}$
at $U/\mu = 0$ as a function of $T/T_{\rm c}$.
This result applies to both the monolayer and bilayer cases.
}
\end{figure}
%%%%%%%%%%%%%%%%%%%%%%%%

\subsubsection{$T$-dependence of the critical current}

The $T$-dependence of the critical current is fully encoded in
eq. (\ref{eq:I_short}).
Here, we estimate this formula numerically,
focusing on the cases of the monolayer and bilayer systems.
The actual computation is performed for
$r = 0.2$, $1.0$, $5.0$, and $r \to \infty$.
The amplitude of the pair potential is determined by the gap equation
\begin{align}
  1 = \lambda_{\rm int} 
  \int_{0}^{\epsilon_{\rm D}} {\rm d}\epsilon
  \tanh\left(\frac{\sqrt{\epsilon^{2}+\Delta^{2}}}{2T}\right)/
  \sqrt{\epsilon^{2}+\Delta^{2}} ,
\end{align}
where $\lambda_{\rm int}$ is the dimensionless interaction constant,
and the Debye energy is chosen as $\epsilon_{\rm D}/\Delta_{0} = 200$.
For comparison we have considered the cases of
homogeneous ($U/\mu = 0$) and highly inhomogeneous ($U/\mu = 10$)
carrier density.

In Fig.~3, the critical current $I_{j{\rm c}}$
(normalized by $I_{0}^{(j)} \equiv e\mathcal{N}_{j}\Delta_{0}$)
is shown as a function of $T/T_{\rm c}$ for the homogeneous case
of $U/\mu = 0$.
Recall that at $U/\mu = 0$, 
the Josephson current for the mono- and bi-layer systems is
described by the single formula of eq.~(\ref{eq:Ij_simplest}).
Thereby, the $T$-dependent normalized critical current for $j=1$ and $j=2$
obviously coincides in this limit.

We observe that $I_{j\rm c}$ is a concave function of $T$
in the strong-coupling limit of $r \to \infty$,
where the KO result is reproduced.
However, it crosses over to a convex function with decreasing $r$.
Such a convex $T$-dependence was not observed
in the previous study~\cite{hagymasi}
based on an energy-independent effective pair potential model.
The coupling strength $\Gamma_{j}$ crucially affects
the $T$-dependence of the critical current.
\begin{figure}[hbtp]
\begin{center}
\includegraphics[height=5.5cm]{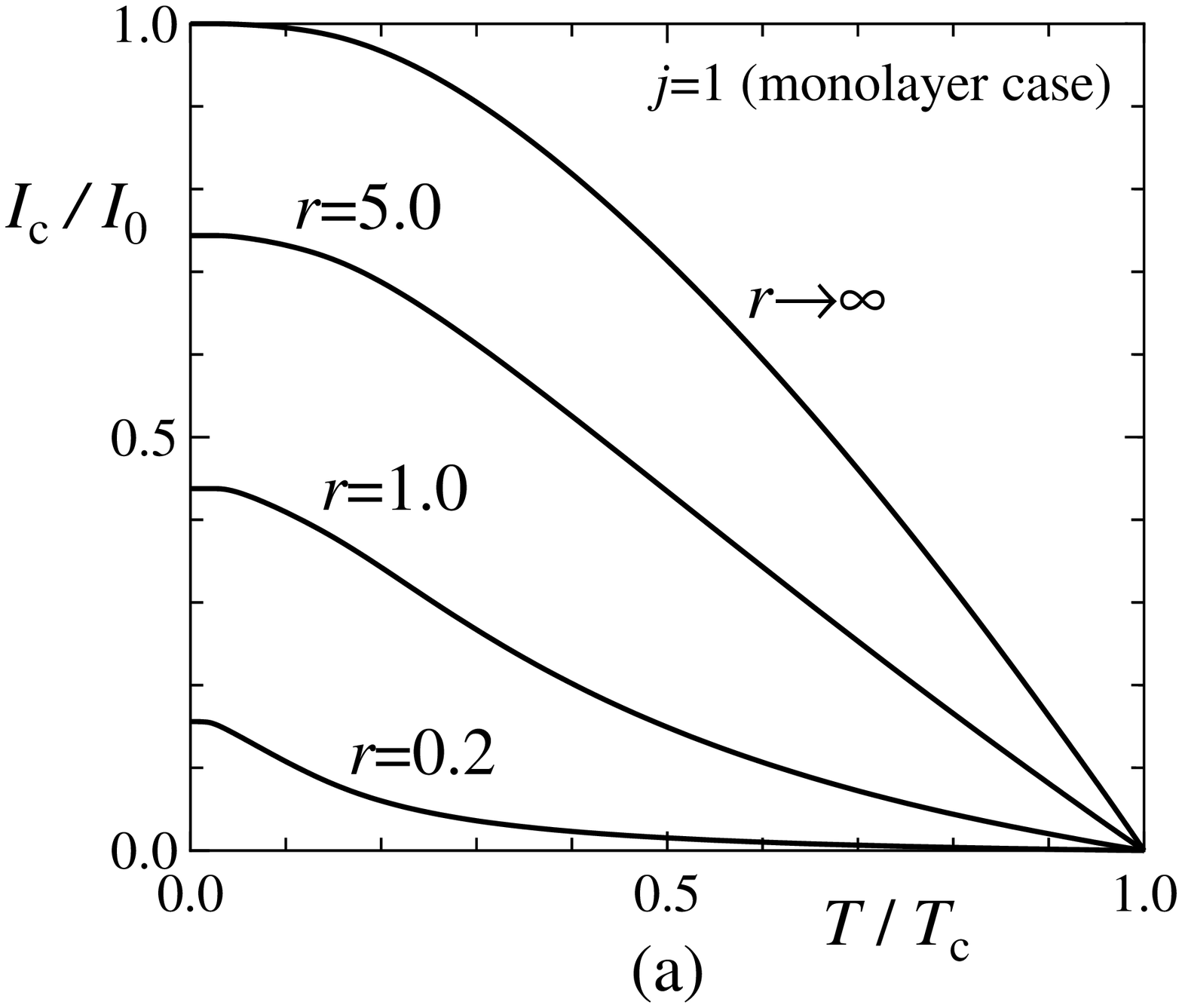}
\includegraphics[height=5.5cm]{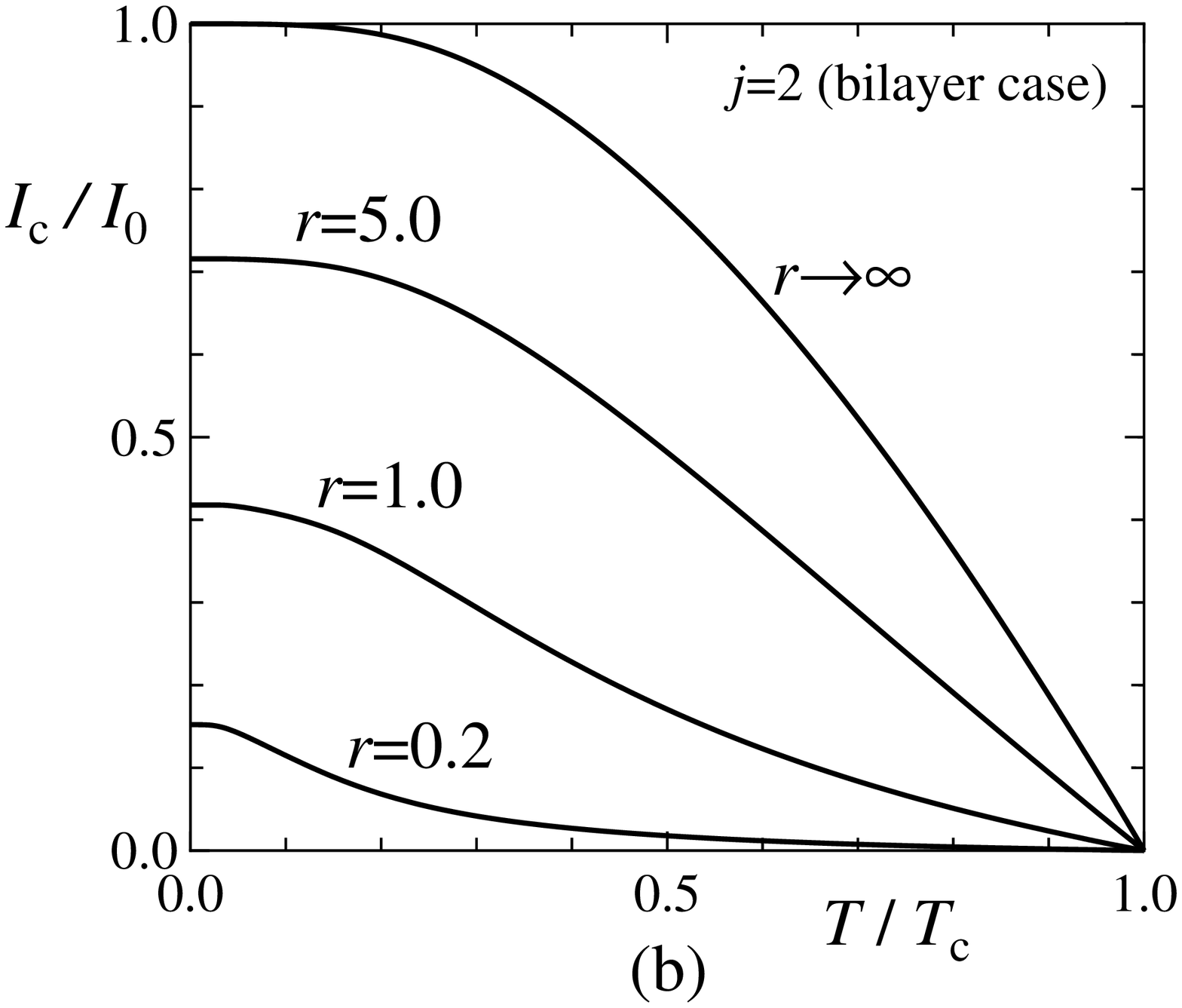}
\end{center}
\caption{The normalized critical current $I_{j{\rm c}}/I_{0}^{(j)}$
at $U/\mu = 10$ for the (a) monolayer and (b) bilayer cases
as a function of $T/T_{\rm c}$.
}
\end{figure}

The behavior of the critical current in the highly inhomogeneous case
of $U/\mu = 10$ is shown in Fig.~4.
Apart from the saturation of the normalized critical current
at low temperatures (and this happens irrespective of the value of $r$),
it seems that an overall character of the $T$-dependence of
the critical current remains unchanged
both in the monolayer and bilayer cases.
This leads us to the following statement:
"the concave-to-convex crossover of the critical current,
which occurs as the decrease of the coupling strength,
which also appears irrespective of the homogeneity of the carrier density,
should be regarded
as a characteristic feature of the planar Josephson junction".

\section{Summary}

After a detailed formulation of our model 
proposed for the planar Josephson junction of graphene,
we have derived and estimated
a class of formulas for the Josephson current.
The derivation was done in the framework of quasiclassical approximation,
and the formulation was generalized to be applicable to the case of
multilayer graphene.
The obtained formulas allow us to estimate the Josephson current numerically
in the planar junction of $N$-layer graphene,
for an arbitrary choice of the set of parameters, that includes
$T$ (temperature), $L$ (separation of two superconductors),
$\Gamma$ (coupling strength between a graphene sheet and superconductors),
and $U$ (effective potential controlling carrier inhomogeneity).
Throughout the paper, we have assumed that
the chemical potential $\mu$ is away from the Dirac point.

Much emphasis has been put on the behavior of the Josephson current
in the short-junction limit in the monolayer and bilayer cases.
It was demonstrated that the coupling strength crucially affects the
temperature dependence of the critical current in an unexpected manner.
This should be regarded as a characteristic feature of the planar junction.
We have also shown that the dependences of the critical current on
$\mu$ and $U$ differ qualitatively
in the monolayer and bilayer cases.
The difference in the $U$-dependence (see Fig. 2) reflects
the contrasting behavior of the transmission probability
across the uncovered-to-covered interface in the two cases.

\section*{Acknowledgment}

Y.T. was supported by a Grant-in-Aid for Scientific Research (C)
(No. 24540375) and K.I. by a Grant-in-Aid for Scientific Research
on Priority Areas ``Topological Quantum Phenomena'' (No. 23103511)
from the Ministry of Education, Culture, Sports, Science and Technology.

\appendix

\section{Derivation of the Boundary Condition}

The unknown coefficients $c_{j++}(\varphi)$ and $c_{j--}(\varphi)$
in the general formula for the Josephson current
[eq.~(\ref{eq:I_start})]
have been fixed by the boundary condition [eq.~(\ref{eq:BC-condition})]
at $x = \pm L/2$.
Here, focusing on the case of $x = L/2$,
we give an explicit derivation of eq.~(\ref{eq:BC-condition}).
The derivation consists of three steps.
Let us start with electron and hole wave functions in the uncovered region
\begin{align}
      \label{eq:Psi_Ne}
  \Psi_{{\rm N}j}^{\rm e}(x;q,\omega)
 & = c_{\rm N+}(x,\omega)\psi_{j+}^{\rm c}(q)
     + c_{\rm N-}(x,\omega)\psi_{j-}^{\rm c}(q) ,
        \\
      \label{eq:Psi_Nh}
  \Psi_{{\rm N}j}^{\rm h}(x;q,\omega)
 & = d_{\rm N+}(x,\omega)\psi_{j+}^{\rm c}(q)
     + d_{\rm N-}(x,\omega)\psi_{j-}^{\rm c}(q) ,
\end{align}
and those in the covered region
\begin{align}
      \label{eq:Psi_Se}
  \Psi_{{\rm S}j}^{\rm e}(x;q,\omega)
 & = c_{\rm S+}(x,\omega)\psi_{j+}^{\rm c}(q)
     + c_{\rm S-}(x,\omega)\psi_{j-}^{\rm c}(q) ,
        \\
      \label{eq:Psi_Sh}
  \Psi_{{\rm S}j}^{\rm h}(x;q,\omega)
 & = d_{\rm S+}(x,\omega)\psi_{j+}^{\rm c}(q)
     + d_{\rm S-}(x,\omega)\psi_{j-}^{\rm c}(q) ,
\end{align}
where $c_{\rm N\pm}(x,\omega)$ and $c_{\rm S\pm}(x,\omega)$ contain
the $x$-dependent factor of ${\rm e}^{\pm{\rm i}k^{+}x}$
and $d_{\rm N\pm}(x,\omega)$ and $d_{\rm S\pm}(x,\omega)$ contain
the factor of ${\rm e}^{\pm{\rm i}k^{-}x}$.
Note that $\psi_{j\pm}^{\rm c}(q)$ and $k$ depend on whether $x$ is
in the covered or uncovered region.
Firstly we relates $c_{\rm S\pm}(L/2,\omega)$
and $d_{\rm S\pm}(L/2,\pm)$ on the basis of
the Bogoliubov-de Gennes equation.
Secondly we derive the relation between
$c_{\rm N\pm}(L/2,\omega)$ and $c_{\rm S\pm}(L/2,\omega)$
and that between $d_{\rm N\pm}(L/2,\omega)$ and $d_{\rm S\pm}(L/2,\omega)$
following the prescription presented by Zaitsev.~\cite{zaitsev}
Finally we derive eq.~(\ref{eq:BC-condition}) by combining these relations.

In order to relate $c_{\rm S\pm}(L/2,\omega)$ and
$d_{\rm S\pm}(L/2,\omega)$, we introduce the Bogoliubov-de Gennes equation
\begin{align}
       \label{eq:BdG-omega}
  \left[ {\rm i}\omega \tau_{2\times2}^{z}
         + {\rm i}\sigma v_{jx} \partial_{x}\tau_{2\times2}^{0}
         - \Sigma_{j}
  \right] \Psi_{{\rm S}j\sigma}(x;q,\omega) = 0 ,
\end{align}
where $\sigma = +$ ($-$) specifies the right-going (left-going) component.
In the region of $x > L/2$, eq.~(\ref{eq:BdG-omega}) is simplified to
\begin{align}
  \left( \begin{array}{cc}
            {\rm i}\tilde{\omega}+{\rm i}\sigma v_{jx}\partial_{x}
            & {\rm i}\tilde{\Delta}{\rm e}^{{\rm i}\varphi/2}  \\
            {\rm i}\tilde{\Delta}{\rm e}^{-{\rm i}\varphi/2}
            & -{\rm i}\tilde{\omega}+{\rm i}\sigma v_{jx}\partial_{x}
         \end{array}
  \right) \Psi_{{\rm S}j\sigma}(x;q,\omega) = 0 .
\end{align}
Its solution is
\begin{align}
  \Psi_{{\rm S}j\sigma}(x;q,\omega)
  = \left( \begin{array}{c}
             \frac{\tilde{\omega}+\sigma\Omega}{\tilde{\Delta}} \\
             {\rm e}^{-{\rm i}\varphi/2}
           \end{array}
    \right) {\rm e}^{-\frac{\Omega}{v_{jx}}(x-L/2)} .
\end{align}
The corresponding four-component solution $\check{\Psi}_{{\rm S}j\sigma}$
is obtained by adding the pseudospin (sublattice) degree of freedom
in accordance with $\psi_{j\sigma}^{\rm c}(q)$, leading to
\begin{align}
  \check{\Psi}_{{\rm S}j\sigma}(x;q,\omega)
  = \left( \begin{array}{c}
             \frac{\tilde{\omega}+\sigma\Omega}{\tilde{\Delta}}
             \psi_{j\sigma}^{\rm c}(q) \\
             {\rm e}^{-{\rm i}\varphi/2} \psi_{j\sigma}^{\rm c}(q)
           \end{array}
    \right) {\rm e}^{-\frac{\Omega}{v_{jx}}(x-L/2)} .
\end{align}
Comparing this with eqs.~(\ref{eq:Psi_Se}) and (\ref{eq:Psi_Sh}),
we easily see that the electron and hole components satisfy
\begin{align}
       \label{eq:e-h_relation}
 &  \left( \begin{array}{c}
              c_{\rm S+}(L/2,\omega) \\
              c_{\rm S-}(L/2,\omega)
           \end{array}
    \right)
      \nonumber \\
 & \hspace{3mm}
  = {\rm e}^{{\rm i}\varphi/2}
    \left( \begin{array}{cc}
              \frac{\tilde{\omega}+\Omega}{\tilde{\Delta}} & 0 \\
              0 & \frac{\tilde{\omega}-\Omega}{\tilde{\Delta}}
            \end{array}
     \right)
    \left( \begin{array}{c}
              d_{\rm S+}(L/2,\omega) \\
              d_{\rm S-}(L/2,\omega)
           \end{array}
    \right) .
\end{align}

We next derive the relation which connects
$c_{{\rm N}\sigma}(L/2,\omega)$ and $c_{{\rm S}\sigma}(L/2,\omega)$,
and $d_{{\rm N}\sigma}(L/2,\omega)$ and $d_{{\rm S}\sigma}(L/2,\omega)$.
Since the effective potential drops to $-U$ across the interface at $x = L/2$
within the length scale comparable to, or shorter than
the Fermi wave length, we are allowed to describe wave functions in
this region by ignoring the influence of $\omega$ and $\Delta$.~\cite{zaitsev}
Hence, with this approximation, we can connect $c_{{\rm N}\sigma}(L/2,\omega)$
and $c_{{\rm S}\sigma}(L/2,\omega)$ by considering a scattering problem
at the Fermi level ignoring the coupling between electron and hole components.
Let $t_{j}$ and $r_{j}$ be the transmission and reflection coefficients
for an electron incident at the interface
from the $x<L/2$-side
with an incident angle $\phi$.
In terms of $t_{j}$ and $r_{j}$
we can construct two independent wave functions:
\begin{align}
   \psi_{1j}(x;q)
 & = \left\{
       \begin{array}{ll}
         \frac{1}{\sqrt{\cos\phi}}
         \big[ c_{\rm N+}(x)\psi_{j+}^{\rm c}(q) \\
                \hspace{11mm}
                + r_{j} c_{\rm N-}(x)\psi_{j-}^{\rm c}(q) \big]
         & (x<L/2) \\
         \frac{1}{\sqrt{\cos\theta}}t_{j}c_{\rm S+}(x)\psi_{j+}^{\rm c}(q)
         & (x>L/2)
       \end{array}
     \right. ,
       \\
   \psi_{2j}(x;q)
 & = \left\{
       \begin{array}{ll}
         \frac{1}{\sqrt{\cos\phi}}
         \tilde{t}_{j}c_{\rm N-}(x)\psi_{j-}^{\rm c}(q)
         & (x<L/2) \\
         \frac{1}{\sqrt{\cos\theta}}
         \big[ c_{\rm S-}(x)\psi_{j-}^{\rm c}(q) \\
                \hspace{11mm}
                + \tilde{r}_{j} c_{\rm S+}(x)\psi_{j+}^{\rm c}(q) \big]
         & (x>L/2)
       \end{array}
     \right. .
\end{align}
Here $\tilde{t}_{j}$ and $\tilde{r}_{j}$ are the transmission and
reflection coefficients for an electron incident from the right,
and they are given by $\tilde{t}_{j} = t_{j}$ and
$\tilde{r}_{j} = -r_{j}^{*}t_{j}/t_{j}^{*}$.
A general wave function is represented by their superposition as
$\psi_{j} = a\psi_{1j} + b\psi_{2j}$.
Comparing this with eqs.~(\ref{eq:Psi_Ne}) and (\ref{eq:Psi_Se}),
we obtain
\begin{align}
     \left( \begin{array}{c}
               c_{\rm N+}(L/2,\omega) \\
               c_{\rm N-}(L/2,\omega)
            \end{array}
     \right)
 & = \frac{1}{\sqrt{\cos\phi}}
     \left(
       \begin{array}{cc}
         1 & 0 \\
         r_{j} & \tilde{t_{j}}
    \end{array}
  \right)
     \left( \begin{array}{c}
               a \\ b
            \end{array}
     \right) ,
           \\
     \left( \begin{array}{c}
               c_{\rm S+}(L/2,\omega) \\
               c_{\rm S-}(L/2,\omega)
            \end{array}
     \right)
 & = \frac{1}{\sqrt{\cos\theta}}
     \left(
       \begin{array}{cc}
         t_{j} & \tilde{r}_{j} \\
         0 & 1
    \end{array}
  \right)
     \left( \begin{array}{c}
               a \\ b
            \end{array}
     \right) .
\end{align}
These two equations yield
\begin{align}
        \label{eq:N-S_relation-e}
     \left( \begin{array}{c}
               c_{\rm N+}(L/2,\omega) \\
               c_{\rm N-}(L/2,\omega)
            \end{array}
     \right)
   = \sqrt{\frac{\cos\theta}{\cos\phi}} M_{j}
     \left( \begin{array}{c}
               c_{\rm S+}(L/2,\omega) \\
               c_{\rm S-}(L/2,\omega)
            \end{array}
     \right)
\end{align}
with
\begin{align}
  M_{j} =
  \left(
    \begin{array}{cc}
      t_{j}^{-1} & (t_{j}^{*})^{-1}r_{j}^{*} \\
      t_{j}^{-1}r_{j} & (t_{j}^{*})^{-1}
    \end{array}
  \right) .
\end{align}
The same relation holds for $d_{\rm N\pm}(L/2,\omega)$ and
$d_{\rm S\pm}(L/2,\omega)$.

Eliminating $c_{\rm S\pm}(L/2,\omega)$ and $d_{\rm S\pm}(L/2,\omega)$
in eq.~(\ref{eq:N-S_relation-e}) and the corresponding relation
for $d_{\rm N\pm}(L/2,\omega)$ and $d_{\rm S\pm}(L/2,\omega)$
with use of eq.~(\ref{eq:e-h_relation}),
we finally arrive at eq.(\ref{eq:BC-condition}).

\section{Transmission Probability across the Interface}

Here, we derive the expressions of the transmission probability
for an electron at the Fermi level being incident at the interface
at $x = L/2$ from the $x<L/2$-side
with an incident angle $\phi$.
For simplicity we shift the origin of the one-dimensional coordinate
to $x = L/2$ and replace $x-L/2 \to x$.

Let us first consider the case of monolayer.
Using the eigenfunctions of $H_{1}$, one can express
the two-component wave function
for this scattering problem as
\begin{align}
        \label{eq:psi1_xq}
   \psi_{1}(x;q)
 & = \left\{
       \begin{array}{ll}
         \frac{{\rm e}^{{\rm i}kx}}{\sqrt{2\cos\phi}}
         \left(\begin{array}{c}
                 {\rm e}^{-{\rm i}\phi/2} \\
                 {\rm e}^{{\rm i}\phi/2}
               \end{array}
         \right) \\
         \hspace{6mm}
       + r_{1}\frac{{\rm e}^{-{\rm i}kx}}{\sqrt{2\cos\phi}}
         \left(\begin{array}{c}
                 {\rm e}^{{\rm i}\phi/2} \\
                 -{\rm e}^{-{\rm i}\phi/2}
               \end{array}
         \right)
         & (x<0) \\
         t_{1}\frac{{\rm e}^{{\rm i}kx}}{\sqrt{2\cos\theta}}
           \left(
             \begin{array}{c}
               {\rm e}^{-{\rm i}\theta/2} \\
               {\rm e}^{{\rm i}\theta/2}
             \end{array}
           \right)
         & (x>0)
       \end{array}
     \right. ,
\end{align}
where the elements of $\psi_{1\pm}^{\rm c}(q)$ is explicitly shown for clarity.
Note that the value of $k \equiv \sqrt{(\tilde{\mu}/\gamma)^{2}-q^{2}}$
differs in the left and right of the interface.
By matching the two components of the wave function at $x = 0$ we obtain
\begin{align}
   t_{1} & = \frac{\sqrt{\cos\phi\cos\theta}}{\cos\frac{\phi+\theta}{2}} ,
         \\
   r_{1} & = {\rm i}
             \frac{\sin\frac{\phi-\theta}{2}}{\cos\frac{\phi+\theta}{2}} .
\end{align}
We easily see that the transmission probability
$\mathcal{T}_{1} \equiv |t_{1}|^{2}$ is given by eq.~(\ref{eq:T1}).

We next consider the bilayer case in which
evanescent modes play a role.~\cite{katsnelson1,katsnelson2}
By using the eigenfunctions of $H_{2}$, the two-component wave function
is expressed as
\begin{align}
      \label{eq:psi2_xq}
   \psi_{2}(x;q)
 & = \left\{
       \begin{array}{ll}
         \frac{{\rm e}^{{\rm i}kx}}{\sqrt{2\cos\phi}}
         \left(\begin{array}{c}
                 {\rm e}^{{\rm i}\phi} \\
                 -{\rm e}^{-{\rm i}\phi}
               \end{array}
         \right) \\
         \hspace{6mm}
       + r_{2}\frac{{\rm e}^{-{\rm i}kx}}{\sqrt{2\cos\phi}}
         \left(\begin{array}{c}
                 {\rm e}^{-{\rm i}\phi} \\
                 -{\rm e}^{{\rm i}\phi}
               \end{array}
         \right) \\
         \hspace{6mm}
       + a{\rm e}^{\kappa x}
         \left(\begin{array}{c}
                 1 \\ \frac{\kappa+q}{\kappa-q}
               \end{array}
         \right)
         & (x<0) \\
         t_{2}\frac{{\rm e}^{{\rm i}kx}}{\sqrt{2\cos\theta}}
         \left(\begin{array}{c}
                 {\rm e}^{{\rm i}\theta} \\
                 -{\rm e}^{-{\rm i}\theta}
               \end{array}
         \right) \\
         \hspace{6mm}
       + b{\rm e}^{-\kappa x}
         \left(\begin{array}{c}
                 1 \\ \frac{\kappa-q}{\kappa+q}
               \end{array}
         \right)
         & (x>0)
       \end{array}
     \right. ,
\end{align}
where $k = \sqrt{(\tilde{\mu}/\alpha)-q^{2}}$, and
$\kappa = \sqrt{(\tilde{\mu}/\alpha)+q^{2}}$.
Note that the value of $k$ and $\kappa$ differ
in the left and right of the interface.
It should be noted that the wave function contains the probability amplitudes
at ${\rm B}_{1}$ and ${\rm A}_{2}$ sites.
Remember that the effective Hamiltonian $H_{2}$ for the bilayer case is
obtained in the framework of a second order perturbation theory.
Within this perturbation theory the probability amplitudes
at ${\rm A}_{1}$ and ${\rm B}_{2}$ sites are obtained from those at
${\rm B}_{1}$ and ${\rm A}_{2}$ sites.
Indeed, for a given two-component wave function
$\psi_{2}(x;q) = \,^{t}(u(x),v(x))$
for ${\rm B}_{1}$ and ${\rm A}_{2}$ sites, we can show that
the corresponding four-component function is given by~\cite{nakanishi}
\begin{align}
 \check{\psi}_{2}(x;q)
 =\,\raisebox{5.5mm}{}^{t}\!
  \left(-\frac{\gamma}{\gamma_{1}}\tilde{k}_{+}v(x),
         u(x),v(x),-\frac{\gamma}{\gamma_{1}}\tilde{k}_{-}u(x)\right) ,
\end{align}
where $\tilde{k}_{\pm} \equiv \hat{k}_{x} \pm {\rm i}q$.
Constructing $\check{\psi}_{2}(x;q)$ from eq.~(\ref{eq:psi2_xq}) and then
matching its each component at $x = 0$,
we finally obtain
\begin{align}
       \label{eq:t2}
   t_{2}
 & = \frac
     {2\left(1+\mathcal{L}\right)\sqrt{w}\sqrt{\cos\phi\cos\theta}}
     {{\rm e}^{-{\rm i}\phi}\mathcal{L}(1+w{\rm e}^{{\rm i}(\phi+\theta)})
      +{\rm e}^{{\rm i}\phi}(1+w){\rm e}^{-{\rm i}(\phi+\theta)}} ,
         \\
   r_{2}
 & = \frac
     {-{\rm e}^{{\rm i}\phi}\mathcal{L}(1-w{\rm e}^{-{\rm i}(\phi-\theta)})
      -{\rm e}^{-{\rm i}\phi}(1-w{\rm e}^{{\rm i}(\phi-\theta)})}
     {{\rm e}^{-{\rm i}\phi}\mathcal{L}(1+w{\rm e}^{{\rm i}(\phi+\theta)})
      +{\rm e}^{{\rm i}\phi}(1+w{\rm e}^{-{\rm i}(\phi+\theta)})} ,
\end{align}
where $w = \sqrt{\mu/(\mu+U)}$ and
\begin{align}
       \label{eq:def-L0}
  \mathcal{L}
  = \frac
    {\left( \sqrt{\frac{\mu}{\alpha}+q^{2}}+q \right)
     \left( \sqrt{\frac{\mu+U}{\alpha}+q^{2}}-q \right)}
    {\left( \sqrt{\frac{\mu}{\alpha}+q^{2}}-q \right)
     \left( \sqrt{\frac{\mu+U}{\alpha}+q^{2}}+q \right)} .
\end{align}
Equations~(\ref{eq:T2}) and (\ref{eq:def-L}) are easily obtained from
eqs.~(\ref{eq:t2}) and (\ref{eq:def-L0}), respectively.

\section{$N$-layer Thermal Green's Function $\check{G}_{N,m}$}

Applying the method presented in ref.~\citen{takane2},
we decompose the $4N \times 4N$ thermal Green's function
$\underline{G}(\mib{r},\mib{r}';\omega)$
into a set of $4 \times 4$ functions.
Let us define
\begin{align}
  \varphi_{0}(i)
  & = \sqrt{\frac{2}{N+1}} (-1)^{i-1} ,
          \\
  \varphi_{m}^{+}(i)
  & =  \frac{2}{\sqrt{{N+1}}} (-1)^{i-1}
       \cos\left(\frac{m\pi}{2(N+1)}(2i-1)\right) ,
          \\
  \varphi_{m}^{-}(i)
  & =  \frac{2}{\sqrt{{N+1}}} (-1)^{i-1}
       \sin\left(\frac{m\pi}{2(N+1)}(2i)\right) ,
\end{align}
for a given $N$.
In terms of these functions we define the following bases
for the odd-$N$ case:
\begin{align}
 |1_{0} \rangle
 & = \varphi_{0}(1)|{\rm A}_{1}\rangle_{e}
     + \varphi_{0}(2)|{\rm A}_{3}\rangle_{e}
     \nonumber \\ & \hspace{15mm}
     + \cdots
     + \varphi_{0}((N+1)/2)|{\rm A}_{N}\rangle_{e} ,
        \\
 |2_{0} \rangle
 & = \varphi_{0}(1)|{\rm B}_{1}\rangle_{e}
     + \varphi_{0}(2)|{\rm B}_{3}\rangle_{e}
     \nonumber \\ & \hspace{15mm}
     + \cdots
     + \varphi_{0}((N+1)/2)|{\rm B}_{N}\rangle_{e} ,
        \\
 |3_{0} \rangle
 & = \varphi_{0}(1)|{\rm A}_{1}\rangle_{h}
     + \varphi_{0}(2)|{\rm A}_{3}\rangle_{h}
     \nonumber \\ & \hspace{15mm}
     + \cdots
     + \varphi_{0}((N+1)/2)|{\rm A}_{N}\rangle_{h} ,
        \\
 |4_{0} \rangle
 & = \varphi_{0}(1)|{\rm B}_{1}\rangle_{h}
     + \varphi_{0}(2)|{\rm B}_{3}\rangle_{h}
     \nonumber \\ & \hspace{15mm}
     + \cdots
     + \varphi_{0}((N+1)/2)|{\rm B}_{N}\rangle_{h} ,
\end{align}
and
\begin{align}
 |1_{m} \rangle
 & = \varphi_{m}^{+}(1)|{\rm A}_{1}\rangle_{e}
     + \varphi_{m}^{+}(2)|{\rm A}_{3}\rangle_{e}
     \nonumber \\ & \hspace{8mm}
     + \cdots
     + \varphi_{m}^{+}((N+1)/2)|{\rm A}_{N}\rangle_{e} ,
        \\
 |2_{m} \rangle
 & = \varphi_{m}^{+}(1)|{\rm B}_{1}\rangle_{e}
     + \varphi_{m}^{+}(2)|{\rm B}_{3}\rangle_{e}
     \nonumber \\ & \hspace{8mm}
     + \cdots
     + \varphi_{m}^{+}((N+1)/2)|{\rm B}_{N}\rangle_{e} ,
        \\
 |3_{m} \rangle
 & = \varphi_{m}^{-}(1)|{\rm A}_{2}\rangle_{e}
     + \varphi_{m}^{-}(2)|{\rm A}_{4}\rangle_{e}
     \nonumber \\ & \hspace{8mm}
     + \cdots
     + \varphi_{m}^{-}((N-1)/2)|{\rm A}_{N-1}\rangle_{e} ,
        \\
 |4_{m} \rangle
 & = \varphi_{m}^{-}(1)|{\rm B}_{2}\rangle_{e}
     + \varphi_{m}^{-}(2)|{\rm B}_{4}\rangle_{e}
     \nonumber \\ & \hspace{8mm}
     + \cdots
     + \varphi_{m}^{-}((N-1)/2)|{\rm B}_{N-1}\rangle_{e} ,
        \\
 |5_{m} \rangle
 & = \varphi_{m}^{+}(1)|{\rm A}_{1}\rangle_{h}
     + \varphi_{m}^{+}(2)|{\rm A}_{3}\rangle_{h}
     \nonumber \\ & \hspace{8mm}
     + \cdots
     + \varphi_{m}^{+}((N+1)/2)|{\rm A}_{N}\rangle_{h} ,
        \\
 |6_{m} \rangle
 & = \varphi_{m}^{+}(1)|{\rm B}_{1}\rangle_{h}
     + \varphi_{m}^{+}(2)|{\rm B}_{3}\rangle_{h}
     \nonumber \\ & \hspace{8mm}
     + \cdots
     + \varphi_{m}^{+}((N+1)/2)|{\rm B}_{N}\rangle_{h} ,
        \\
 |7_{m} \rangle
 & = \varphi_{m}^{-}(1)|{\rm A}_{2}\rangle_{h}
     + \varphi_{m}^{-}(2)|{\rm A}_{4}\rangle_{h}
     \nonumber \\ & \hspace{8mm}
     + \cdots
     + \varphi_{m}^{-}((N-1)/2)|{\rm A}_{N-1}\rangle_{h} ,
        \\
 |8_{m} \rangle
 & = \varphi_{m}^{-}(1)|{\rm B}_{2}\rangle_{h}
     + \varphi_{m}^{-}(2)|{\rm B}_{4}\rangle_{h}
     \nonumber \\ & \hspace{8mm}
     + \cdots
     + \varphi_{m}^{-}((N-1)/2)|{\rm B}_{N-1}\rangle_{h} .
\end{align}
In the odd-$N$ case, we can decompose the $4N \times 4N$ Hamiltonian
$\underline{H}$ into one monolayer-type Hamiltonian with dimensions
$4 \times 4$ and $(N-1)/2$ bilayer-type Hamiltonians with dimensions
$8 \times 8$ by transforming the original bases to
the new ones presented just above.~\cite{takane2}.
Applying this procedure, the $4N \times 4N$
thermal Green's function $\underline{G}(\mib{r},\mib{r}';\omega)$
is decomposed into one monolayer-type Green's function
$\check{G}_{N,0}(\mib{r},\mib{r}';\omega)$
and $(N-1)/2$ bilayer-type Green's functions
$\widehat{G}_{N,m}(\mib{r},\mib{r}';\omega)$
with $m = 2,4,\dots, N-1$.
In doing so we approximately ignore weak coupling
between different Green's functions due to the self-energy.~\cite{takane2}
The $4 \times 4$ monolayer-type Green's function $\check{G}_{N,0}$ is defined
so that its $(i,j)$-element ($i,j = 1,2,3,4$) is given by
\begin{align}
  \left[\check{G}_{N,0}\right]_{i,j}
   = \langle i_{0}|\underline{G}| j_{0} \rangle .
\end{align}
From eq.~(\ref{eq:Green-N}) we can show that this obeys eq.~(\ref{eq:eq-G}).
The $8 \times 8$ bilayer-type Green's function $\widehat{G}_{N,m}$ is defined
so that its $(i,j)$-element ($i,j = 1, 2, 3, \dots, 8$) is given by
\begin{align}
  \left[\widehat{G}_{N,m}\right]_{i,j}
   = \langle {i}_{m}|\underline{G}| {j}_{m} \rangle .
\end{align}
From eq.~(\ref{eq:Green-N}) we can show that this obeys
\begin{align}
  \left( {\rm i}\omega \widehat{\tau}_{8\times8}^{z}-\widehat{H}_{N,m}
          -\widehat{\Sigma}_{N,m}
  \right) \widehat{G}_{N,m}(\mib{r},\mib{r}';\omega)
  = \widehat{\tau}_{8\times8}^{0}\delta(\mib{r}-\mib{r}')
\end{align}
with $\widehat{\tau}_{8\times8}^{z}
= \tau_{2\times2}^{z}\otimes\check{\tau}_{4\times4}^{0}$,
$\widehat{\tau}_{8\times8}^{0}
= \tau_{2\times2}^{0}\otimes\check{\tau}_{4\times4}^{0}$, and
$\widehat{H}_{N,m} = {\rm diag}(\check{H}_{2}^{N,m}, \check{H}_{2}^{N,m})$.
The Hamiltonian $\check{H}_{2}^{N,m}$ is
\begin{align}
  \check{H}_{2}^{N,m}
  = \left( \begin{array}{cc}
             H_{1} & \lambda_{N,m}V \\
             \lambda_{N,m}V^{\dagger} & H_{1}
           \end{array}
    \right) 
\end{align}
with $\lambda_{N,m}$ given in eq.~(\ref{eq:def-lambda}).
The self-energy is
\begin{align}
  \widehat{\Sigma}_{N,m}
& = \frac{-2{\rm i}\Gamma_{N,m}}{\sqrt{\Delta^{2}+\omega^{2}}}
    \left( \begin{array}{cc}
             \omega & \Delta(x) \\
             \Delta(x)^{*} & -\omega
           \end{array}
    \right)
    \nonumber \\
&   \hspace{5mm} \otimes
    \left( \begin{array}{cc}
             \tau_{2\times2}^{0} & 0_{2\times2} \\
             0_{2\times2} & 0_{2\times2}
           \end{array}
    \right)
    \theta\left(|x|-\frac{L}{2}\right) ,
\end{align}
where $\Gamma_{N,m} \equiv (1/2)\varphi_{m}^{+}(1)^{2}\Gamma$.
Now we reduce the $8\times8$ Green's function $\widehat{G}_{N,m}$ to
a $4\times4$ Green's function in accordance with
the argument presented by MacCann and Fal'ko.~\cite{maccann}
The interlayer coupling (i.e., $\gamma_{1}$ in $V$ and $V^{\dagger}$)
forms dimers in the electron space spanned by $|1_{m} \rangle$
and $|4_{m} \rangle$, and in the hole space spanned by
$|5_{m} \rangle$ and $|8_{m} \rangle$.
Since the energy of dimer states is greater than $\gamma_{1}$,
we can safely ignore these states as long as $\tilde{\mu} \ll \gamma_{1}$.
Consequently, within a second order perturbation theory,
low-energy quasiparticles are described
in the electron-hole space spanned by
$|2_{m} \rangle$, $|3_{m} \rangle$, $|6_{m} \rangle$, $|7_{m} \rangle$
and the corresponding $4\times4$ Hamiltonian is given by
$\check{H}_{N,m}={\rm diag}(H_{2}^{N,m}, H_{2}^{N,m})$
with the reduced bilayer-type Hamiltonian $H_{2}^{N,m}$
presented in eq.~(\ref{eq:H_2-Nm}).
In accordance with this reduction, $\widehat{G}_{N,m}$ is reduced to
the $4\times4$ Green's function $\check{G}_{N,m}$ which obeys
eq.~(\ref{eq:eq-G}).

In the even-$N$ case, the $4N \times 4N$ thermal Green's function
is decomposed into $N/2$ bilayer-type Green's functions
$\widehat{G}_{N,m}(\mib{r},\mib{r}';\omega)$ with $m = 1,3,\dots, N-1$.
The bilayer-type Green's function
$\widehat{G}_{N,m}(\mib{r},\mib{r}';\omega)$ is defined
so that its $(i,j)$-element ($i,j = 1, 2, 3,\dots,8$) is given by
\begin{align}
  \left[\widehat{G}_{N,m}\right]_{i,j}
   = \langle {i}_{m}|\underline{G}| {j}_{m} \rangle
\end{align}
with
\begin{align}
 |1_{m} \rangle
 & = \varphi_{m}^{+}(1)|{\rm A}_{1}\rangle_{e}
     + \varphi_{m}^{+}(2)|{\rm A}_{3}\rangle_{e}
     \nonumber \\ & \hspace{15mm}
     + \cdots
     + \varphi_{m}^{+}(N/2)|{\rm A}_{N-1}\rangle_{e} ,
        \\
 |2_{m} \rangle
 & = \varphi_{m}^{+}(1)|{\rm B}_{1}\rangle_{e}
     + \varphi_{m}^{+}(2)|{\rm B}_{3}\rangle_{e}
     \nonumber \\ & \hspace{15mm}
     + \cdots
     + \varphi_{m}^{+}(N/2)|{\rm B}_{N-1}\rangle_{e} ,
        \\
 |3_{m} \rangle
 & = \varphi_{m}^{-}(1)|{\rm A}_{2}\rangle_{e}
     + \varphi_{m}^{-}(2)|{\rm A}_{4}\rangle_{e}
     \nonumber \\ & \hspace{15mm}
     + \cdots
     + \varphi_{m}^{-}(N/2)|{\rm A}_{N}\rangle_{e} ,
        \\
 |4_{m} \rangle
 & = \varphi_{m}^{-}(1)|{\rm B}_{2}\rangle_{e}
     + \varphi_{m}^{-}(2)|{\rm B}_{4}\rangle_{e}
     \nonumber \\ & \hspace{15mm}
     + \cdots
     + \varphi_{m}^{-}(N/2)|{\rm B}_{N}\rangle_{e} ,
        \\
 |5_{m} \rangle
 & = \varphi_{m}^{+}(1)|{\rm A}_{1}\rangle_{h}
     + \varphi_{m}^{+}(2)|{\rm A}_{3}\rangle_{h}
     \nonumber \\ & \hspace{15mm}
     + \cdots
     + \varphi_{m}^{+}(N/2)|{\rm A}_{N-1}\rangle_{h} ,
        \\
 |6_{m} \rangle
 & = \varphi_{m}^{+}(1)|{\rm B}_{1}\rangle_{h}
     + \varphi_{m}^{+}(2)|{\rm B}_{3}\rangle_{h}
     \nonumber \\ & \hspace{15mm}
     + \cdots
     + \varphi_{m}^{+}(N/2)|{\rm B}_{N-1}\rangle_{h} ,
        \\
 |7_{m} \rangle
 & = \varphi_{m}^{-}(1)|{\rm A}_{2}\rangle_{h}
     + \varphi_{m}^{-}(2)|{\rm A}_{4}\rangle_{h}
     \nonumber \\ & \hspace{15mm}
     + \cdots
     + \varphi_{m}^{-}(N/2)|{\rm A}_{N}\rangle_{h} ,
        \\
 |8_{m} \rangle
 & = \varphi_{m}^{-}(1)|{\rm B}_{2}\rangle_{h}
     + \varphi_{m}^{-}(2)|{\rm B}_{4}\rangle_{h}
     \nonumber \\ & \hspace{15mm}
     + \cdots
     + \varphi_{m}^{-}(N/2)|{\rm B}_{N}\rangle_{h} .
\end{align}
Repeating the treatment similar to that presented in the odd-$N$ case,
we can reduce $\widehat{G}_{N,m}$ into $\check{G}_{N,m}$ which again obeys
eq.~(\ref{eq:eq-G}).


\begin{thebibliography}{99}
\bibitem{comment1}
Here we have two comments to be made on the effective potential.
Firstly, if the characteristic length scale of this potential is
much longer than the Fermi wave length,
we can neglect its influence on the Josephson current
since such a smooth potential does not induce electron scattering.
Secondly, the assumption of a constant $-U$ breaks down if the thickness of
multilayer graphene becomes much larger than the inter-layer screening length.
Such a case is beyond scope of this paper,
and deserves separate consideration.
\bibitem{comment2} $\sqrt{E_{+}(\varphi)E_{-}(\varphi)}$
in eq.~(31) of ref.~\citen{takane1} should read $E_{+}(\varphi)E_{-}(\varphi)$.
\end{thebibliography}
\end{document}